\pdfoutput=1
\newlength\smallfigwidth
\documentclass[aip,jap,preprint,floatfix,showpacs,amsmath,amssymb]{revtex4-1}

\usepackage{color}
\usepackage{graphicx}
\def\ba{\begin{eqnarray}}
\def\ea{\end{eqnarray}}
\def\be{\begin{equation}}
\def\ee{\end{equation}}

\begin{document}

\title{Magnetization ground state and reversal modes of magnetic nanotori}
\author{Smiljan Vojkovic}
\affiliation{Instituto de F\'isica, Facultad de F\'isica, Pontificia Universidad Cat\'olica de Chile, \\Campus San Joaqu\'in - Av. Vicu\~na Mackenna 4860, Santiago, Chile.}
\author{\'Alvaro S. N\'u\~nez}
\affiliation{Departamento de F\'isica, Facultad de Ciencias F\'isicas y 
Matem\'aticas, Universidad de Chile, \\Casilla 487-3, Santiago, Chile}
\author{Dora Altbir}
\affiliation{Departamento de F\'isica, Universidad de Santiago de Chile and CEDENNA,\\ Avda. Ecuador 3493, Santiago, Chile}
\author{Vagson L. Carvalho-Santos}
\email{vagson.santos@ufv.br}
\affiliation{Instituto Federal de Educa\c c\~ao, Ci\^encia e Tecnologia Baiano - Campus 
Senhor do Bonfim, \\Km 04 Estrada da Igara, 48970-000 Senhor do Bonfim, Bahia, Brazil}

\date{\today}

\begin{abstract}

In this work, and by means of micromagnetic simulations, we study the magnetic properties of toroidal nanomagnets. The magnetization ground state for different values of the aspect ratio between the toroidal and polar radii of the nanotorus has been obtained. The hysteresis curves are also obtained, evidencing  the existence of two reversal modes depending on the geometry: a vortex mode  and a coherent rotation. 
A comparison between toroidal and cylindrical nanoparticles has been performed  evidencing that  nanotori can accommodate a vortex as the ground state for smaller volume than cylindrical nanorings.  This is  important  because if vortices are used as bits of information, nanotori allow a higher density data storage.
\end{abstract}

\maketitle
\section{Introduction}
During the last decades, nanomagnetism has received strong attention due to advances in the controlled production of magnetic nanoparticles with several shapes and sizes \cite{Cowburn-JPD-33-2000,Fabric-1,Fabric-2,Fabric-3,Fabric-4,Fabric-5}. The increasing interest in this area lies on the possibility of using magnetic nanostructures  in diverse applications  such as  data storage and  random access memory devices \cite{Possibilities} as well as cancer therapy \cite{Cancer-therapy}, among others. For such applications, but also from the fundamental point of view,  it is  important to describe and understand the several aspects  behind the static and dynamic properties of the magnetization. Among the factors that can influence the  behavior of the magnetization,  geometry plays an important role \cite{Cowburn-JPD-33-2000}. In this context, magnetic curved structures such as spheres \cite{Goll-PRB-2004,Russier-JAP-2009,Kravchuk-PRB-2012}, cylinders \cite{Landeros-APL-2007} and cones \cite{Cone-paper} have been frequently studied focusing on  the influence of the curvature on these structures. From the theoretical point of view, this interest has been reinforced due to the recent development of a functional that allows to calculate the exchange energy of nanomagnets with arbitrary shapes \cite{Gaididei-PRL-2014,Sheka-JPA-2014}. This functional has been used to study the magnetic properties of non-planar nanomagnets such as M\"obius stripes \cite{Kravchuk-Mobius}, helical wires \cite{Kravchuk-Helical} and domain walls in a parabolic local bend of a nanowire \cite{Kravchuk-curved-nanowire}. 

Recently it has been shown that curvature can induce changes in the energy of vortices and skyrmions in a paraboloidal nanoshell, introducing a new characteristic length into the system. This characteristic length is responsible for a shrunk of the skyrmion to small regions of the nanomagnet. Besides, the exchange energy of vortices is larger in curved regions of the nanoparticle \cite{Vagson-JAP-2015,Priscila-PLA-2015}. In addition, some theoretical works have pointed out the possibility of controlling the magnetic state of a nanoparticle by  coupling a magnetic field to its curvature, inducing the formation of $2\pi$-skyrmions on cylindrically-shaped magnetic shells \cite{Vagson-JMMM-2015}.  

\begin{figure}
\includegraphics[scale=0.25]{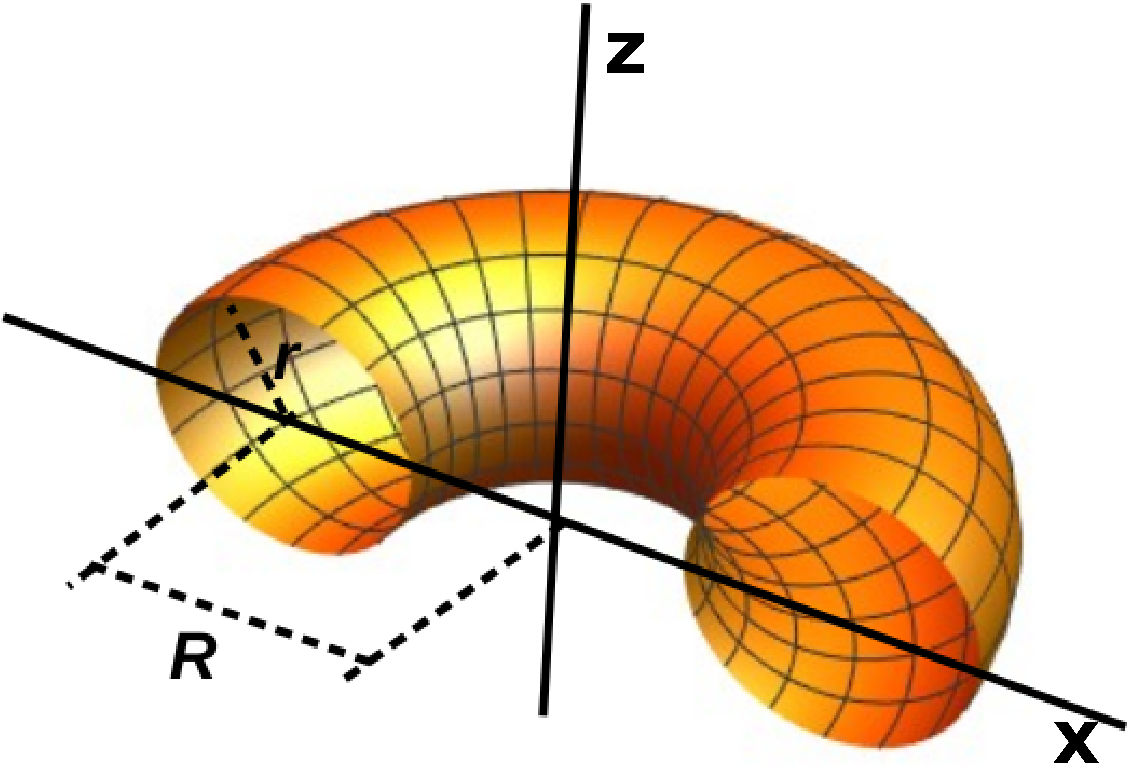}\\\includegraphics[scale=0.1]{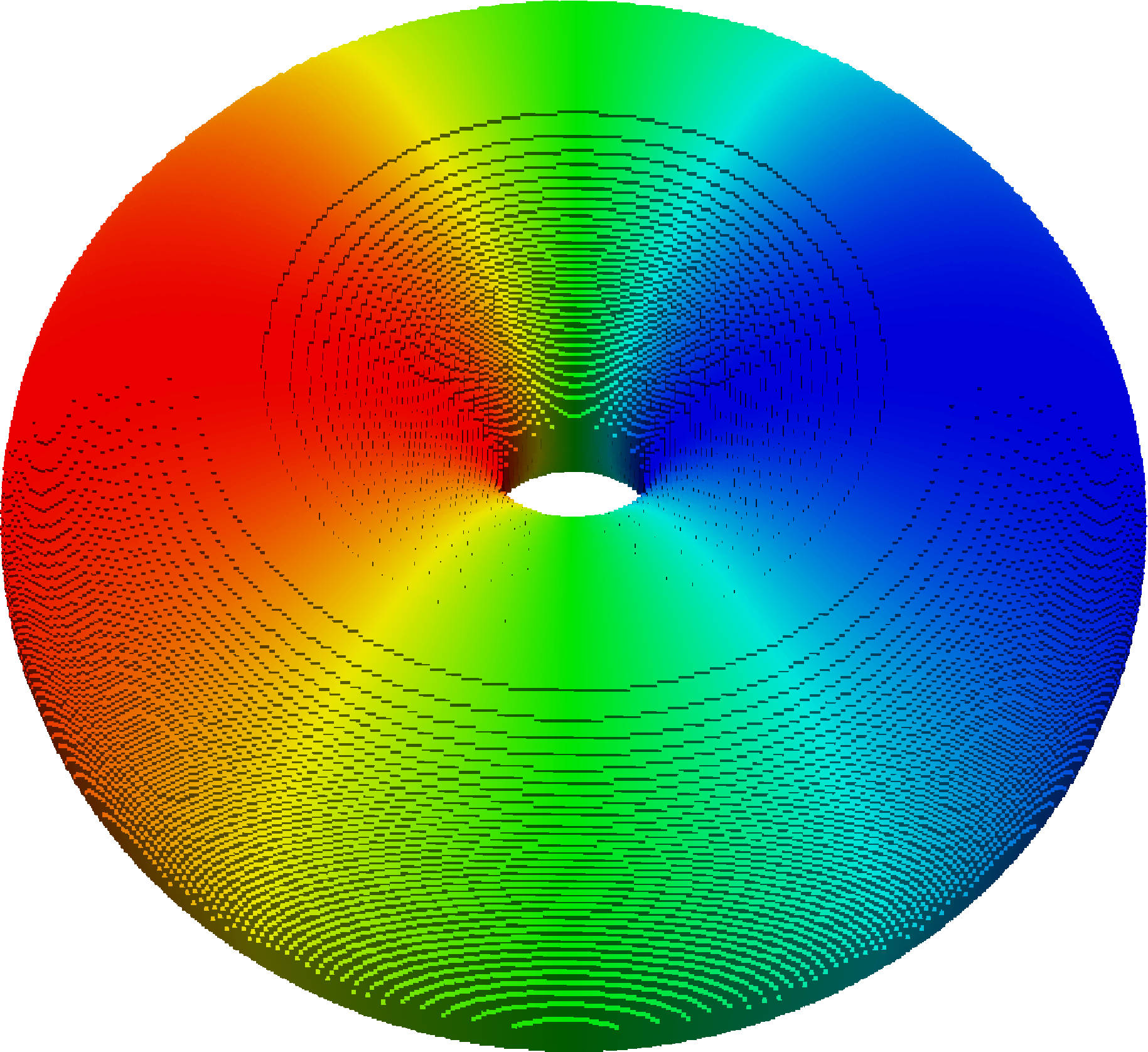}\caption{Top figure illustrates the geometric parameters describing a torus. The $xz$-plane is represented together with $R$ and $r$ defined as the toroidal and polar radii, respectively. Bottom figure shows a simulated toroidal structures, with $R=14$ nm and $r=11$ nm, evidencing that the discretization produces a nanoparticle with smooth surface.}\label{Half-Torus}
\end{figure}

A particular and interesting geometry largely studied is a cylindrical nanoring \cite{Klaui-works,Yoo-APL-2003,Heyderman-JAP-2003,Klaui-JAP-2004,Klaui-APL-2004,Bellegia-JMMM-2006,Ross-JAP-2006,Yang-APL-2007,Zhu-PRL-2006,Vaz-JPC-2007,Kravchuck-nanoring,Castillo-works}. One remarkable feature of such nanorings is the appearance of vortices as the magnetic ground state for smaller sizes than cylindrical nanodots due to the ring hole, which avoids the formation of the core, diminishing the dipolar energy cost to maintain a magnetic vortex \cite{Kravchuck-nanoring}. Moreover, the switching mechanism of the vortex circulation depends on the aspect ratio between the external and internal radii of the nanoring \cite{Klaui-works,Castillo-works}. In this context, it is  interesting to understand how the magnetic properties of a small nanoring with the geometry and topology of a torus change when compared to cylindrical rings. Although cylindrical rings share the torus topology, its geometry does not, once at the disk and hole borders, curvature abruptly changes. In addition, the evaluation of magnetostatic energies, which are very sensitive to the size and geometry of the magnet, seems to be more manageable for the toroidal geometry due to its smoother curvature. Nevertheless, toroidal nanomagnets have been scarcely explored in the literature. In part, this could be associated with the fact that despite the fabrication of magnetic nanoparticles with different geometries is already possible \cite{Fabric-1,Fabric-2,Fabric-3,Fabric-4,Fabric-5}, to produce magnetic nanoparticles with toroidal shape is still a hard task. However,  there have been efforts to prepare nanotori in non-magnetic materials. For instance, carbon nanotori \cite {Carbon-nanotori} and polymer microtori \cite{polymer-tori} have been produced and studied as potential candidates for contention vessels for applications in nanomagnetism, nanobionic and nanobiometric devices. Thus, the preparation of magnetic nanotori may soon be a reality. From the theoretical point of view, it has been shown that nanomagnets with this shape support a vortex state for smaller radius than their cylindrical counterparts \cite{Vagson-JAP-2010} and a systematic study about dipolar energy of single domain states on toroidal magnets has been performed \cite{Bellegia-Torus}. Moreover, due to the appearance of two characteristic lengths associated to the torus geometry (toroidal and polar radii), it has been shown that a geometrical frustration  leads to a skyrmion instability \cite{Dandoloff-torus,Vagson-PRB-2008}.

Based on the above, we study the magnetization ground state of ferromagnetic nanotori by performing micromagnetic simulations using the 3D OOMMF code \cite{oommf-code} for different aspect ratios $R/r$, with $R$ and $r$  the toroidal and polar radii of the toroidal nanomagnet, respectively (see Fig. \ref{Half-Torus}). A previous analytical study   addressed some issues in such structures\cite{Vagson-JAP-2010}, finding the geometrical parameters for which a  vortex configuration represents  the magnetization ground state in toroidal nanomagnets. In this work we go deeper into this topic by showing that the obtained phase diagram slightly change due to small deviations in the magnetization that are present  in the single domain state. We have also compared the minimum volumes of nanotori and cylindrical nanorings for which a vortex appears as the magnetization ground state, searching for higher density data storage. Furthermore, we obtain hysteresis curves of  the magnetic nanotori and identify two different reversal modes, a  coherent rotation $(\mathcal{CR})$ and vortex $(\mathcal{V})$ reversal modes.

\section{Formalism and methods}

For our calculations we use  micromagnetism, a model that neglects the atomistic character of the systems for the description of the static and dynamics of the magnetization, which is defined as the density of magnetic moments. Under its frame, a magnetic structure is described as a continuous medium whose magnetic state is defined by the magnetization vector as a function of the position inside the element. The physical foundation of micromagnetics was laid by Landau, Lifshitz \cite{Landau-paper} and Gilbert \cite{Gilbert-paper}, who  developed a phenomenological equation describing the time evolution of the magnetization in a ferromagnet. In this model the dynamics of the magnetization is determined by the torque acting on the magnetic moment of each volume element,  due to an effective field $H_{\text{eff}}$. The equation, known as the Landau-Lifshitz-Gilbert equation, is given by
\begin{figure}
\includegraphics[scale=0.15]{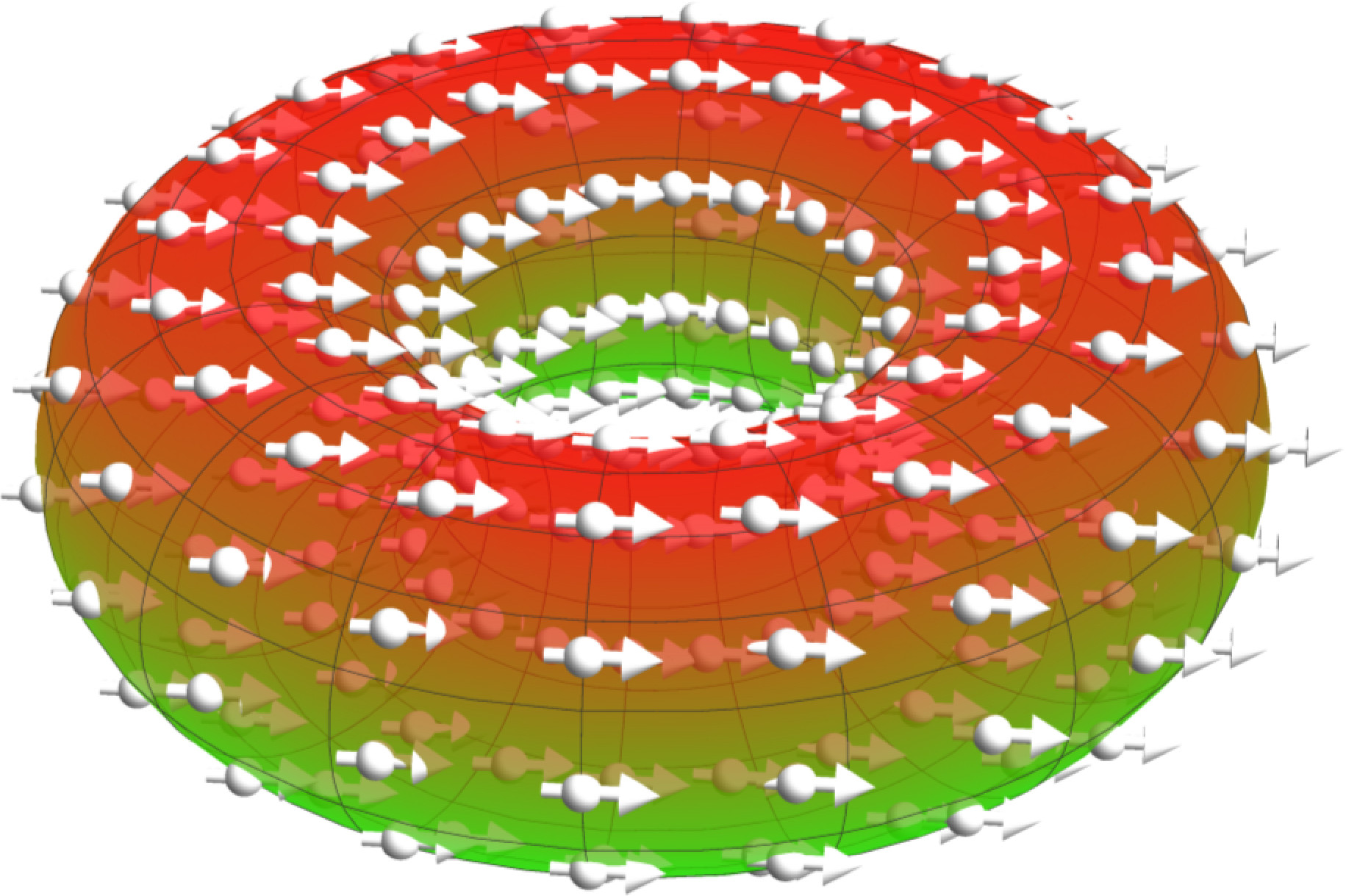}\includegraphics[scale=0.143]{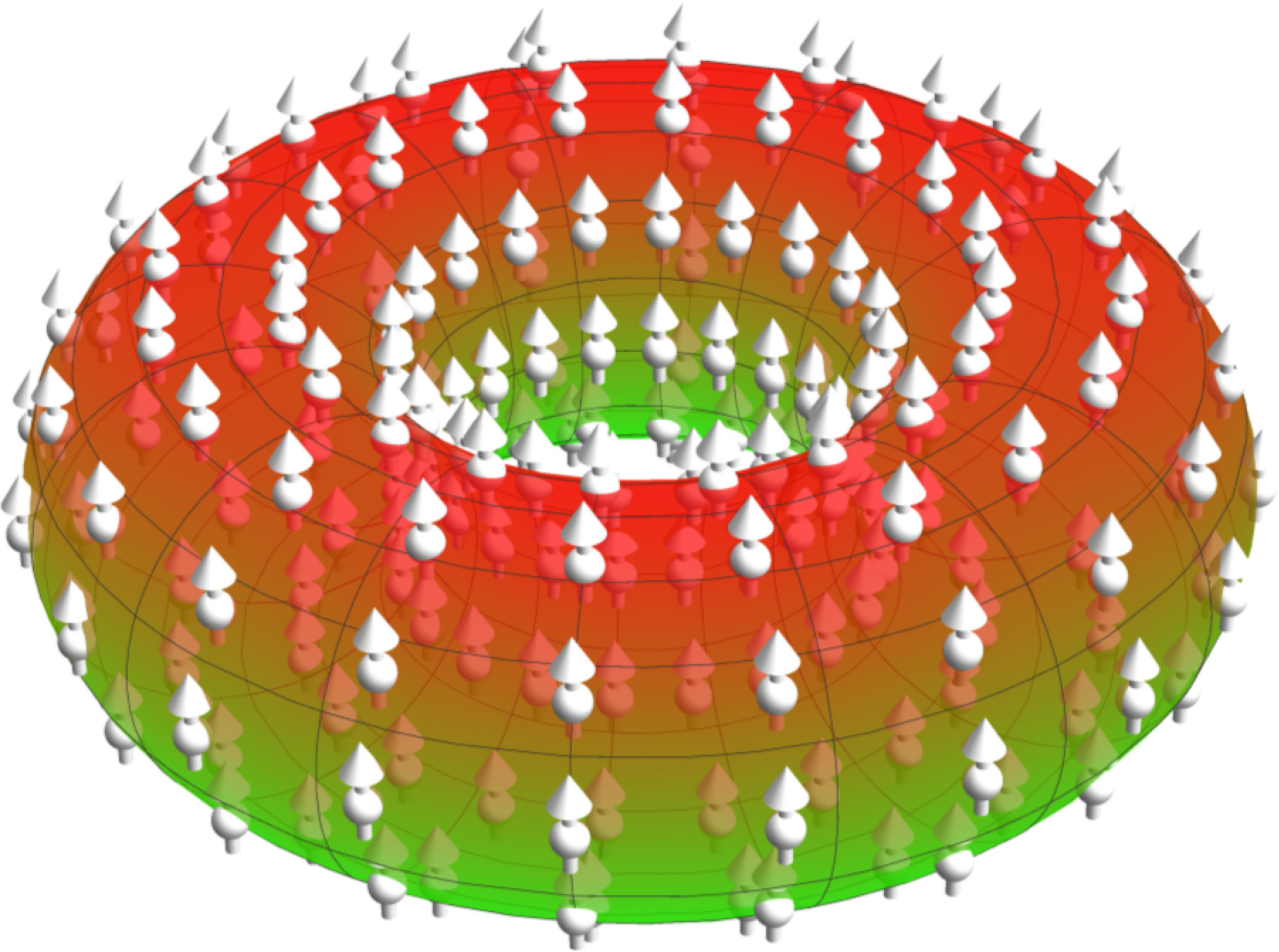}\includegraphics[scale=0.175]{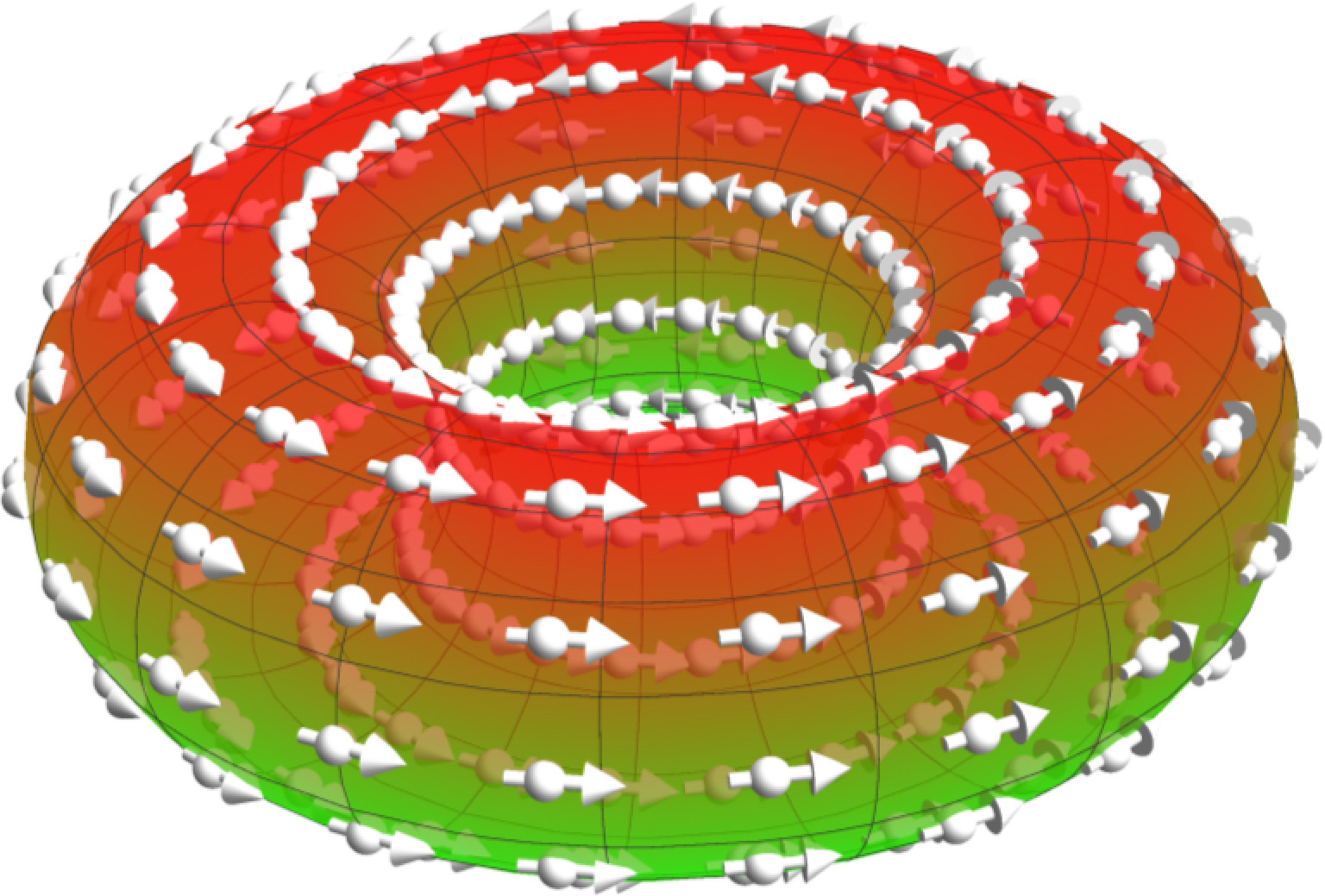}\caption{Initial  configurations of the magnetization from micromagnetic simulations. From left to right we show the $\text{SD}_\text{xy}$, $\text{SD}_\text{z}$ and the in-plane vortex configurations.}\label{Initial-Configurations}
\end{figure}

\be\label{LLGEq}
\frac{\partial \mathbf{m}}{\partial t}=\frac{\gamma_0}{\mu_0M}\mathbf{m}\times\frac{\partial\mathcal{E}}{\partial\mathbf{m}}+\alpha\mathbf{m}\times\frac{\partial\mathbf{m}}{\partial t}\,,
\ee
where $\gamma_0=\mu_0|\gamma|=\mu_0g|\mu_B|/\hbar$, with $\gamma$ the gyromagnetic ratio, $\mathbf{m}=\mathbf{M}/{M_{_S}}$ is the magnetization direction with $M_{_S}$ the saturation magnetization, $\mathcal{E}$ is the free energy density and $\alpha$ is the dimensionless Gilbert damping parameter. While the first term of Eq. (\ref{LLGEq}) describes the precession of the magnetization under the influence of an effective magnetic field $\mathbf{H}_{\text{eff}}=(-1/\mu_0M_{_S})\delta\mathcal{E}/\delta\mathbf{m}$, the second one accounts for the relaxation mechanisms that dissipate energy by making a torque towards the effective field. The effective magnetic field that each magnetic moment experiments is created by the exchange, dipolar and anisotropy interactions, as well as the external magnetic field. In our calculations for simplicity we will consider no anisotropy.

We consider permalloy nanotori of different aspect ratios $R/r$,  and solve Eq. (\ref{LLGEq})  using the 3D  Object Oriented MicroMagnetic Framework (OOMMF) package \cite{oommf-code}. The simulations were run using an exchange constant $A=1.3\times10^{-11}$ J/m, a saturation magnetization $M_{_S}=860$ A/m, and a mesh size of  $0.2\times0.2\times0.2$ nm$^3$ which ensures that the magnetic structure has the toroidal shape even when we consider a small polar radius. For instance, nanotori with $r=1$ nm are 10 cells height and the transversal area of their arms have approximately 78 cells. In Fig. \ref{Half-Torus} we show a nanotorus with $R=14$ nm and $r=11$ nm composed by 4,071,252 cells, showing that this discretization generates geometric objects that are not a superposition of concentric rings with variable radii but very similar to an ideal torus. The damping constant  is assumed zero, however, aiming to allow the system reaches equilibrium, we have used a torque condition that establishes when to stop the simulations. That is,  the minimum energy state is reached when the torque on the magnetic moments is below $0.01$ A/m.

Our simulations were performed starting three different initial configurations: a single domain along the $z$-axis direction  ($\text{SD}_\text{z}$), a single domain along  the $xy$-plane direction ($\text{SD}_\text{xy}$) and a vortex state ($\text{V}$), illustrated in Fig. \ref{Initial-Configurations}. We have left the system relax until  it reaches the minimum energy, as explained above.  Using the calculated energies we have obtained a phase diagram for magnetic nanotori with  $r$ varying from 1 nm to 13 nm and $R$ from $r+1$ nm to 14 nm.  Finally, we have performed micromagnetic simulations  to identify the reversal mechanisms of the magnetization for two specific directions of the magnetic field, $z$-axis and $xy$-plane. For the hysteresis calculations and  to reduce computational times  we have increased the cell size to $1\times1\times1$ nm$^3$ and the damping constant to $0.5$. However we have tested that this change does not induces significant quantitative changes with respect to results obtained with the initial parameters.
\begin{figure}
\includegraphics[scale=0.5]{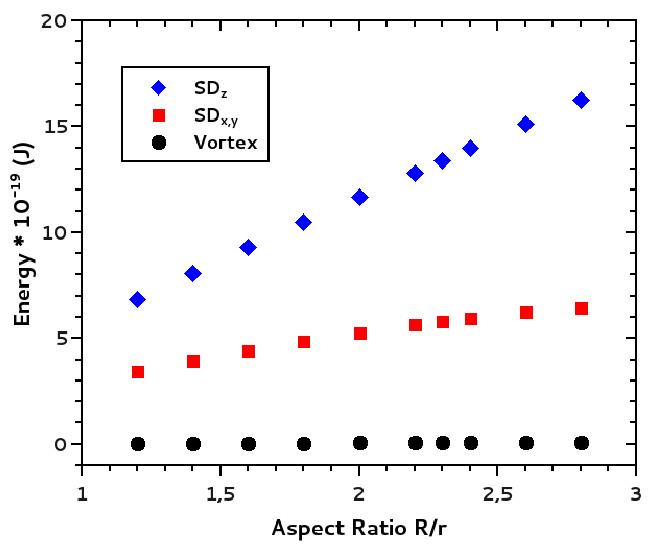}\caption{Dipolar energy of toroidal structures after relaxing from from each of the three initial configurations of the magnetization, for $r=5$ nm.}\label{DipEnFig}
\end{figure}

\begin{figure}
\includegraphics[scale=0.3]{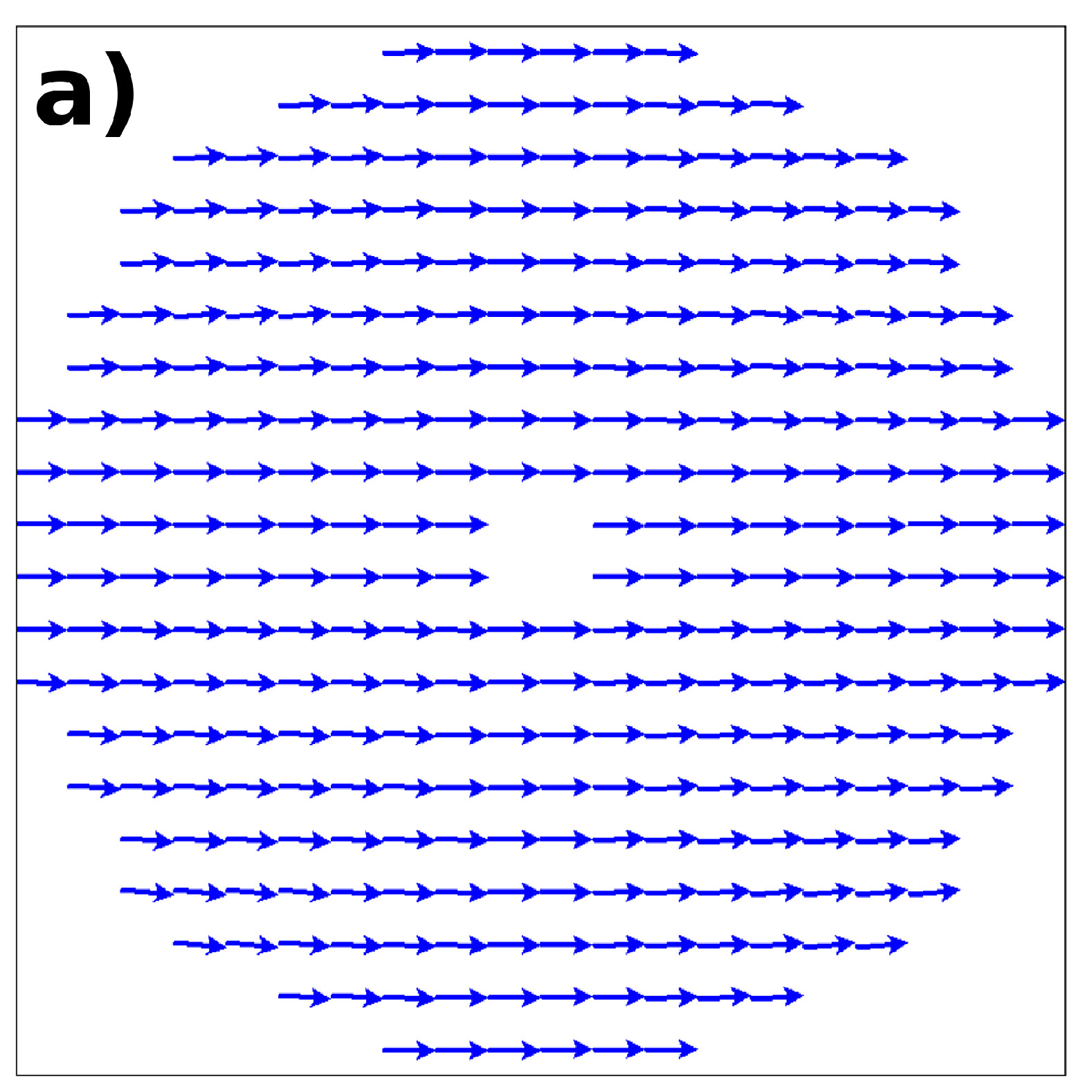}\hspace{0.1em}\includegraphics[scale=0.3]{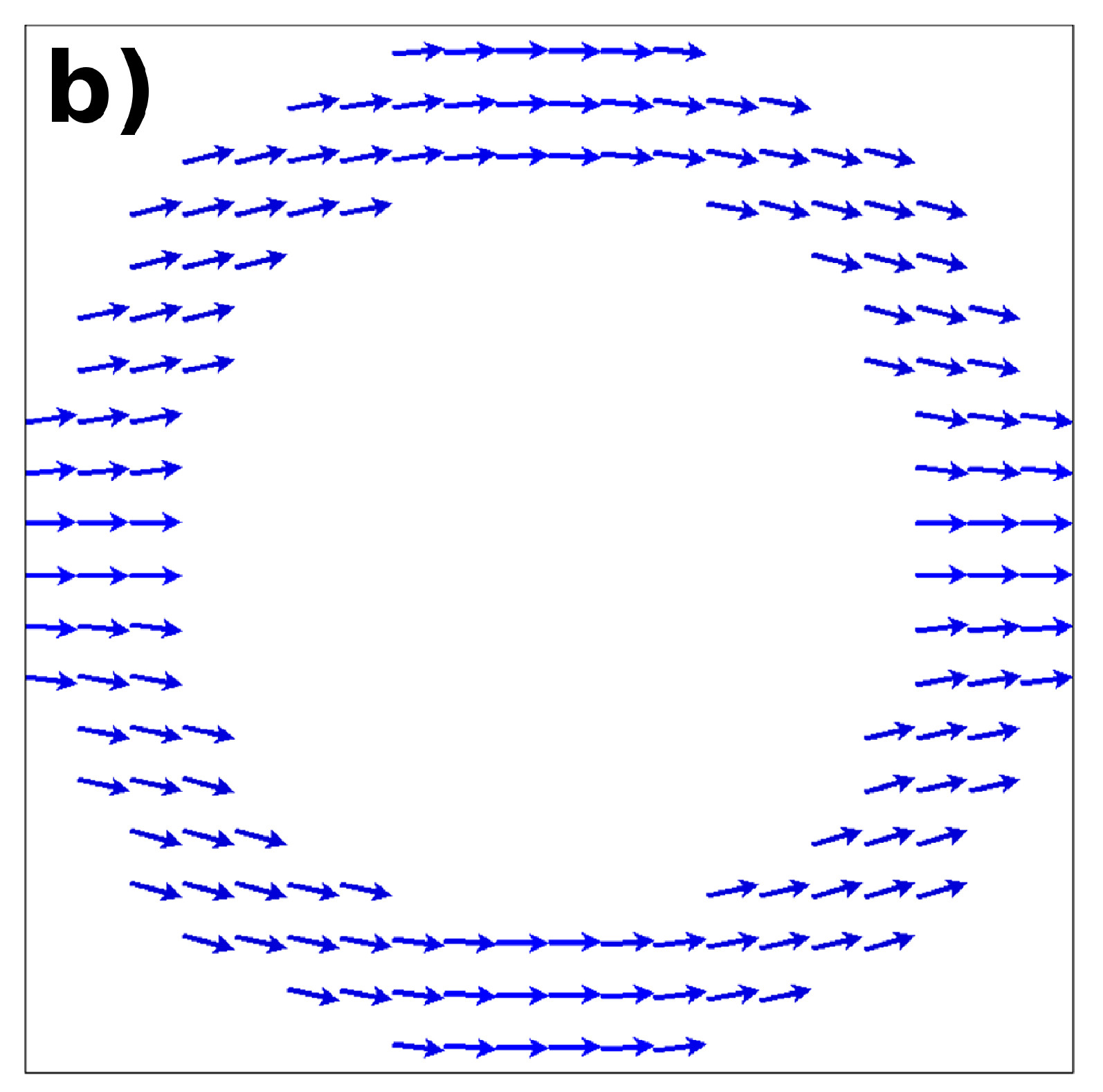}\caption{Illustration of the $\text{SD}_\text{xy}$ state after relaxation for $r=6$ nm and (a) $R=12$ and (b) $26$ nm, respectively. One can note the presence of small deviations from the purely parallel configuration..}\label{Relax-SD-Vortex}
\end{figure}

\section{Magnetic energy and phase diagram}

We proceed by  obtaining  the dipolar energy associated to the lower energy states obtained starting from  each of the three  initial configurations presented above, $\text{SD}_\text{z}$, $\text{SD}_\text{xy}$, and $\text{V}$, as a  function of the aspect ratio. The results for  $r=5$ nm are shown in Fig. \ref{DipEnFig}. Since in the  $\text{SD}_\text{z}$  and in the  $\text{SD}_\text{xy}$ states the exchange energy is the same, the occurrence of one or the other as the lower energy state depends only on the dipolar energy contribution. Fig. \ref{DipEnFig} evidences that the dipolar energy of the $\text{SD}_\text{z}$ state is around twice the  dipolar energy  associated to the $\text{SD}_\text{xy}$ configuration, in agreement with results presented in Ref. [37]. Thus, a toroidal nanomagnet will never  assume a $\text{SD}_\text{z}$ ordering, and therefore, we will not consider it from now on. Despite the dipolar energy of the $\text{SD}_\text{z}$ state varies linearly with the volume of the nanomagnet, the magnetostatic energy of the $\text{SD}_\text{xy}$ configuration does not follow this linear behavior. Small deviations from the parallel state of the magnetization, the so called onion state, ensures the minimization of the energy and the torque condition (see Fig. \ref{Relax-SD-Vortex}b). In this way there is a smooth decreasing in the dipolar energy associated to the $\text{SD}_\text{xy}$ configuration. An analytical model describing  these small deviations of the magnetic moments from the parallel direction can be found in Refs. [37] and [44].

\begin{figure}
\includegraphics[scale=0.5]{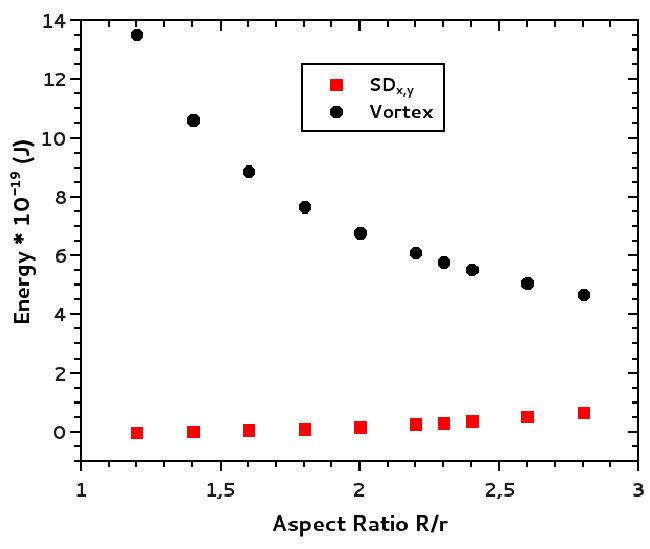}\caption{Exchange energy for the $\text{SD}_{xy}$ and vortex configurations for $r=5$ nm. We can note that while the exchange energy associated to the $\text{SD}_{\text{xy}}$ state is practically zero, the exchange energy of the vortex state diverges for small $R/r$ and tends to zero for large $R/r$. }\label{ExEnFig}
\end{figure}

\begin{figure}
\includegraphics[scale=0.5]{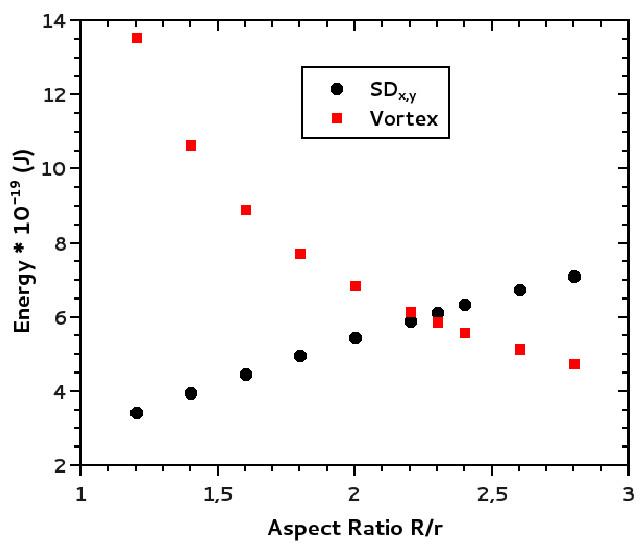}\caption{ Total energy associated to the $\text{SD}_{xy}$ and vortex configurations for $r=5$ nm.}\label{TotEnFig}
\end{figure}

Another interesting point  from the analysis of Fig. \ref{DipEnFig} is that the vortex state presents a very small but finite dipolar energy. This non-zero value  obeys to  the appearance of small magnetic charges associated to the discretization of the toroidal nanomagnet. Nevertheless, once the shell size is tiny, this dipolar cost is very small.

The exchange energy associated to each configuration has been also obtained. Since there is a deviation from the fully saturated  state, the exchange energy associated to the $\text{SD}_\text{xy}$ state is not zero (see Fig. \ref{ExEnFig}).  Nevertheless, the values of the exchange energy of the final state is very small, associated to the fact that the deviations of the magnetic moments from the saturated in-plane state are small (see Fig. \ref{Relax-SD-Vortex}). On the other hand, an increasing in the exchange energy  for large $R$ is observed,  showing that the larger $R/r$ the larger  the deviation of the magnetization from the parallel state. Besides, and as expected,  the exchange energy of the vortex configuration diverges when the radius of the hole in the torus is very small ($R/r\rightarrow1$)  because in this situation in the internal border of the torus hole the direction of the magnetization of neighbor magnetic moments differ appreciably. For the opposite reason, for large values of $R$ the exchange energy associated to the vortex configuration decreases.

\begin{figure}
\includegraphics[width=8.5cm]{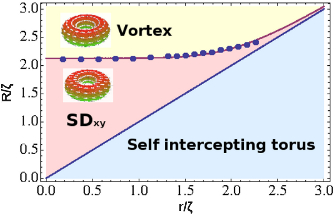}\caption{Phase diagram for nanotori. The white region (yellow on-line)  denotes the phase space  in which the vortex state minimizes the magnetic energy. The grey region (pink on-line) denotes the phase space at which the SD$_{\text{xy}}$ is the lower energy state. The solid line represents $R = r$, that is the geometrical limits of a tori. The light grey region corresponds to a physically non-realizable torus}\label{State-Diagram}
\end{figure}

The total energy of the two contributions, dipolar plus exchange terms, is depicted in Fig. \ref{TotEnFig}  for $r=5$ nm allowing  to obtain the stable states of the magnetization for  nanotori of different aspect ratios. The crossing of the energies occurs  for $R/r\sim2.3$, showing that for smaller values of the aspect ratio the total energy of the  $\text{SD}_\text{xy}$ configuration is smaller, while tori with larger aspect ratios, that is, structures with  small arm thickness, present  vortices as lower energy configuration.

We have performed equivalent simulations for several values of $r$ and obtained the phase diagram illustrated in Fig. \ref{State-Diagram}. This diagram shows that the aspect ratio at which the transition $\text{SD}_{\text{xy}}$ to vortex occurs diminishes when $r$ increases. That is, the greater $r$ the smaller the aspect ratio at which the vortex state becomes stable.  The phase diagram  brings the data  using  two dimensionless quantities $\xi=R/\zeta$ and $\rho=r/\zeta$, where $\zeta=5.3$ nm is the exchange length of permalloy. The  results obtained from the micromagnetic simulations are represented by dots that indicate the transition  separating both minimum-energy states. The line passing through these dots is a fit obtained from the formula 
\begin{eqnarray}
\xi(\rho)=\left(\rho^\alpha+\beta\right)^{1/\alpha} ,
\end{eqnarray}
where $\alpha$ and $\beta$ are adjustable parameters. We used  $\alpha=6.548$ and $\beta=140.333$.

\begin{figure}
\includegraphics[scale=0.3]{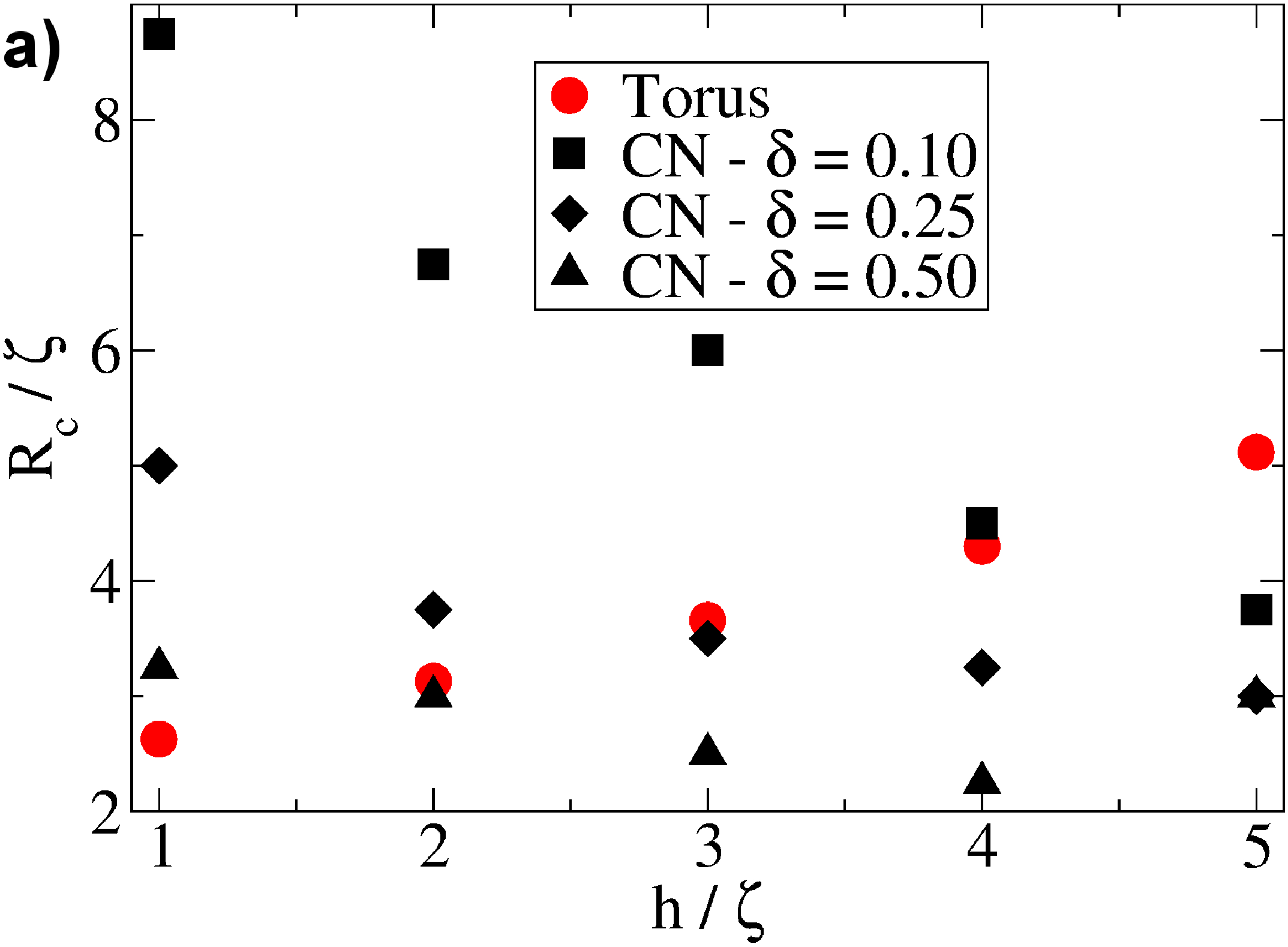}\\\vspace{1em}\includegraphics[scale=0.3]{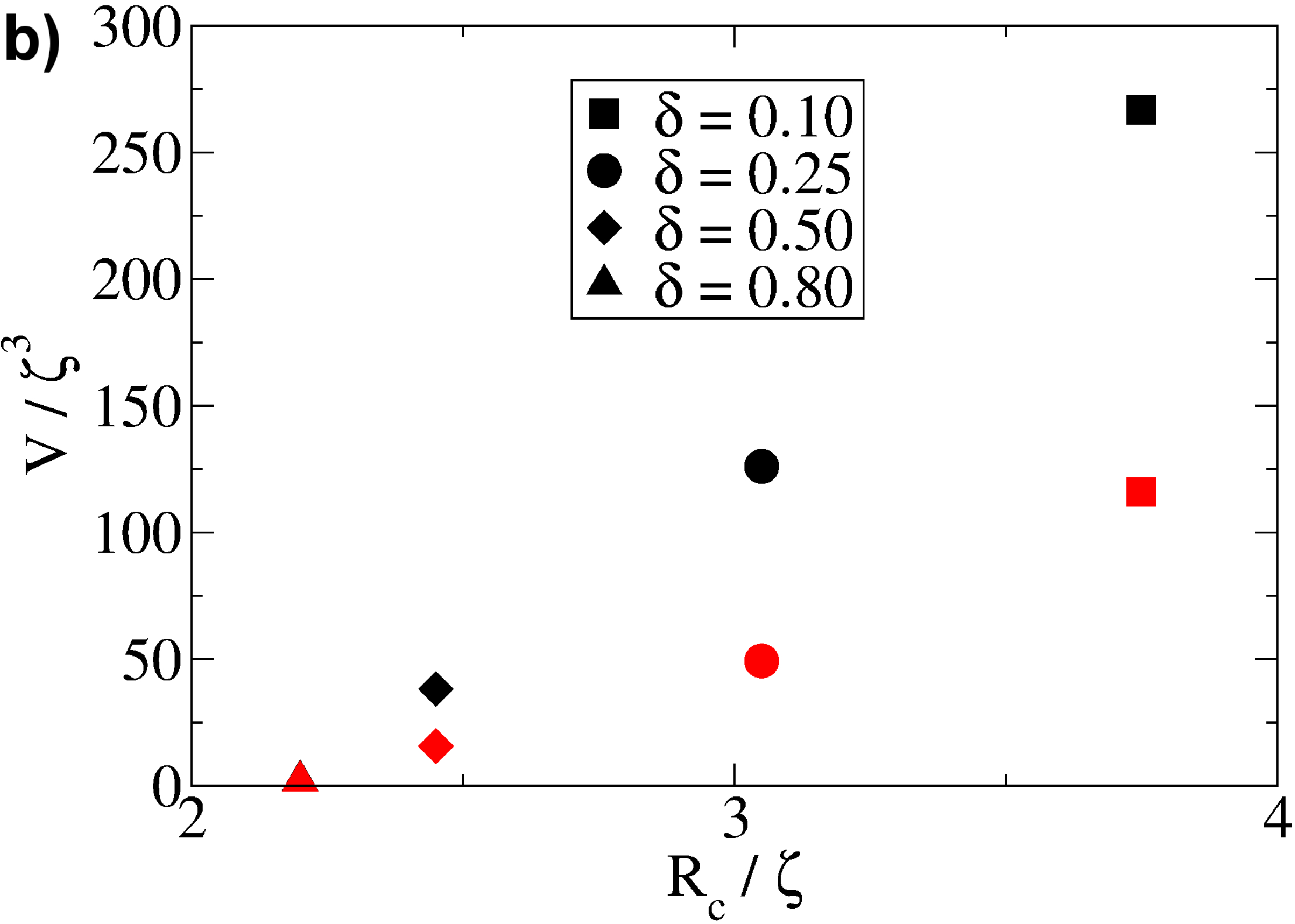}
\caption{(a)  Minimum external radius as a function of the height for toroidal and cylindrical nanorings (CN)  that exhibit a vortex as lower energy state.   (b) Illustration of the corresponding volume as a function of the external radius. Dark symbols represent CN while grey (red) symbols represent nanotori. In this case, toroidal nanoparticles have always lower volume than CN.}\label{Comparison}
\end{figure}

From the analysis of the phase diagram one can note that thick torus with $r/\zeta\lesssim2$ presents  $\text{SD}_{\text{xy}}$ states. But for $r/\zeta\geq2$ the region in which this configuration minimizes the energy diminishes in a such way that even when $R/r\approx1$,  the lower energy configuration is the vortex. On the other hand, for thin magnetic torus, $R/r\gtrsim 10$, the vortex is the  ground state. This phase diagram is very similar to the one presented in  Fig. 5 of Ref. [37], obtained from analytical calculations. However, a comparison between both shows that the  $\text{SD}_{\text{xy}}$ region  is slightly shifted in Fig. \ref{State-Diagram}, generating  a larger area where $\text{SD}_{\text{xy}}$ is the magnetization ground state. This shift in the phase state can be explained by the small deviations from the fully saturated configuration occurring in the magnetization due to the curvature of the nanomagnet. That is, when we  relax the system starting from a $\text{SD}_{\text{xy}}$ state, the final configuration is an onion state, whose magnetic energy is lower than the energy associated to the purely saturated configuration considered in the theoretical model of Ref. [37]. In this way,  results given in Fig. \ref{State-Diagram} offer a better description of the magnetization ground state of a ferromagnetic nanotorus. The self-intercepting torus region in Fig. \ref{State-Diagram} represents  a torus with $R/r<1$,  a physically non-relizable structure.

An intriguing aspect of the magnetization configuration of a permalloy nanotorus is that when $r/\zeta\gtrsim2.5$, the vortex state is stable for any aspect ratio $R/r\approx1.1$. Therefore we can state that the vortex state is stable for any nanotorus whose diameter is larger than 40 nm, independent of its aspect ratio. On the other hand, if $r/\zeta\lesssim1$,  the vortex state appears  only when $R/\zeta>2$. 

Since the torus and the cylindrical ring share the same topology, we have investigated the dimensions at which the vortex becomes the ground state for nanotorus and cylindrical nanorings (CN). A CN can be described by three parameters, height $h_{\text{c}}$, external $R_{\text{c}}$ and internal radii $r_{\text{c}}$. On the other hand, the height of the torus depends on the internal radius  in the form $h_{\text{t}}=2r$, and then only two parameters are needed to describe the toroidal geometry. The parameters of a CN and a nanotorus are related by $R_\text{c}=R+r$ and $r_{c}=R-r$ and their volumes are $V_{\text{c}}=\pi h_{\text{c}}R_c^2(1-\delta^2)$ and $V_{\text{t}}=\pi^2Rh_{\text{t}}^2/2$, where $\delta=r_c/R_c$ is the ratio between the internal and external radii of the CN. To calculate the external radius and volume at which the vortex state is stable in cylindrical rings, we have used the data from Refs. [33] and [44], where the phase diagrams  for magnetic nanorings with radii   $\delta=$0.1, 0.25, 0.5 \cite{Kravchuck-nanoring} and 0.8 \cite{Landeros-JAP-2006} have been presented.

Fig. \ref{Comparison}a illustrates the minimum external radius for nanotori and cylindrical rings at which we obtain a vortex, for different ring $\delta$ values. It can be notice that for small heights, the torus presents a vortex for smaller radius than the CN. However, when the height increases the  external radius needed to generate a vortex in the torus is larger. This behavior is due to the intrinsic relation in nanotori between $h_{\text{t}}$ and $r$, in such a way that the increasing of the height of the torus  implies an enlargement of the polar radius. It can be also noted that the height for which the CN presents a lower external radius than a nanotorus depends on $\delta$. For instance, thick CN ($\delta=0.1$) presents an external radius  lower than that one corresponding to a nanotorus when $h/\zeta\approx4.1$. 

Aiming to compare nanoparticles with similar sizes, we have considered the cases in which tori and  CNs have the same internal and external radii. That is, given a CN characterized by a particular  $\delta$ with the smaller possible external radius that can accommodates a  vortex,  we compare its volume  with the one of a  nanotorus with toroidal radius $R=R_{\text{c}}-r$ and $\delta=(R-r)/(R+r)$. Fig. \ref{Comparison}b presents the  volumes for both structures as a function of their external radii. The volumes of the CNs have been calculated from available data \cite{Kravchuck-nanoring,Landeros-JAP-2006} and are given by dark symbols, while nanotori volumes are given by light symbols. It can be be noticed that toroidal structures always present lower volumes than CNs. The difference between the volumes is larger for small $\delta$ values, with nanotori having less than a half of the CN volume for $\delta=0.1$ and $\delta=0.25$. Therefore, magnetic nanotori can accommodate stable vortices for smaller volumes than CNs with the same external dimensions.

\section{Hysteresis curves and reversal modes}

\begin{figure}
\includegraphics[scale=0.3]{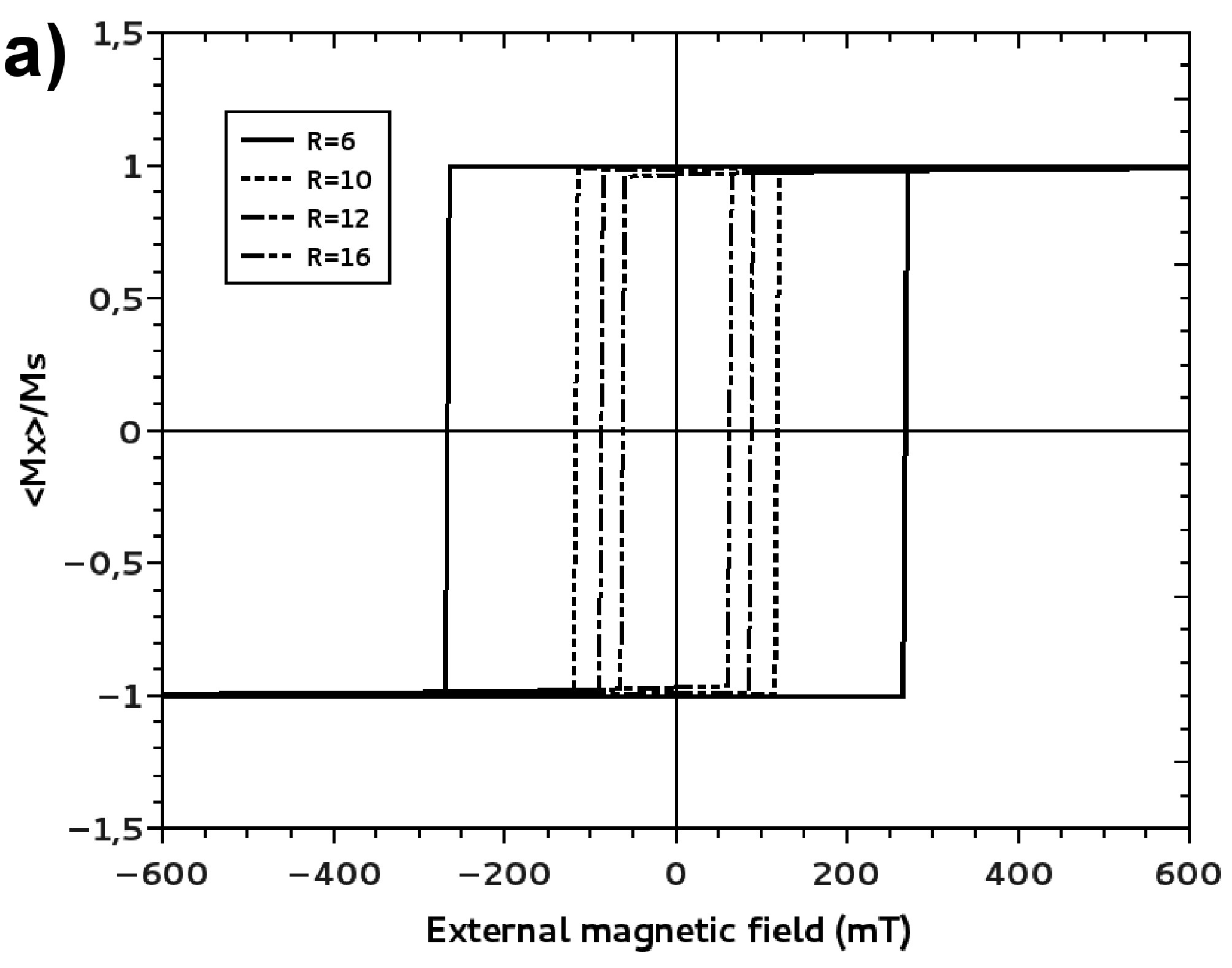}\includegraphics[scale=0.3]{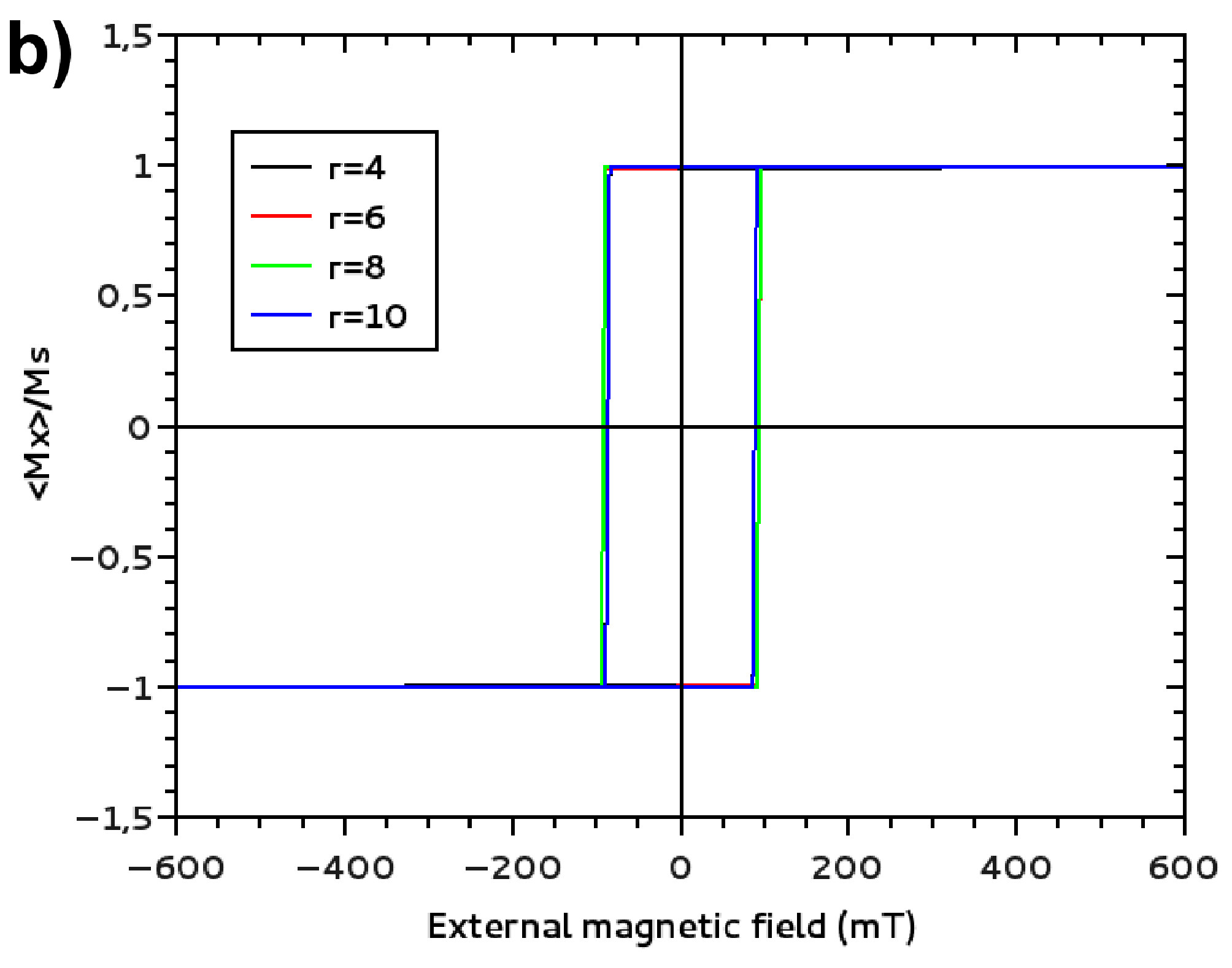}\caption{a) Hysteresis loops for magnetic nanotori with $r=4$ nm and small $R$ under an in-plane magnetic field. b) Hysteresis loops for $R =12$nm and different values of $r$.}\label{Hysteresis-r4-1}
\end{figure}

\begin{figure}
\includegraphics[scale=1]{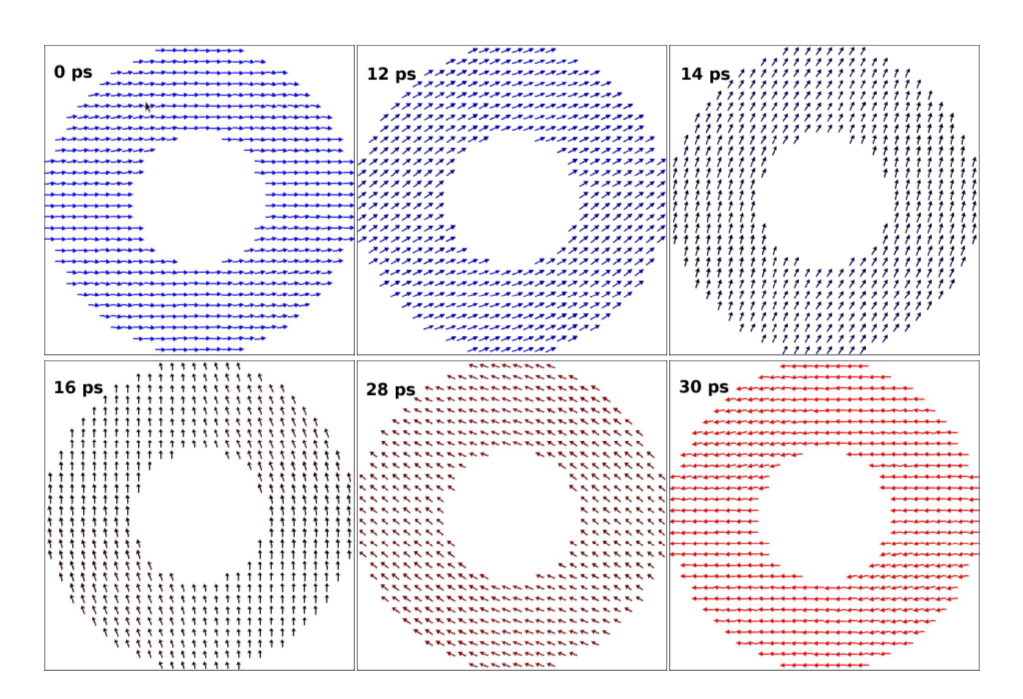}
\caption{Snapshots of the magnetization reversal process of a torus with $R=10$ nm and $r=4$ nm. In this case, the magnetic moments reverse in a fast process at $H_x =$ -120 mT.}\label{Fig-x-fat}
\end{figure}

After determining the lower energy configurations we focus on the magnetization reversal processes by looking to the hysteresis curves. Two directions of the external magnetic field are considered, an in-plane external magnetic field $H_{\text{x}}$ and an out-of-plane external field $H_\text{z}$. For these situations we look into the metastable states appearing during the magnetization reversal process. We start obtaining the hysteresis curves under an in-plane magnetic field  acting on the system using the  OOMMF code. In Fig. 9a we present the hysteresis loops for nanotori with $r=4$ nm and toroidal radius from $6$ nm to $16$ nm. Fig. 9b depicts the hysteresis loops of fix $R = 12$ nm and $r$ from 4 to 10 nm. All loops exhibit a square shape with a coercivity that strongly depends only on $R$,  as  evidenced by the overlaps of the hysteresis cycles in Fig. 9b. 

\begin{figure}
\includegraphics[scale=0.3]{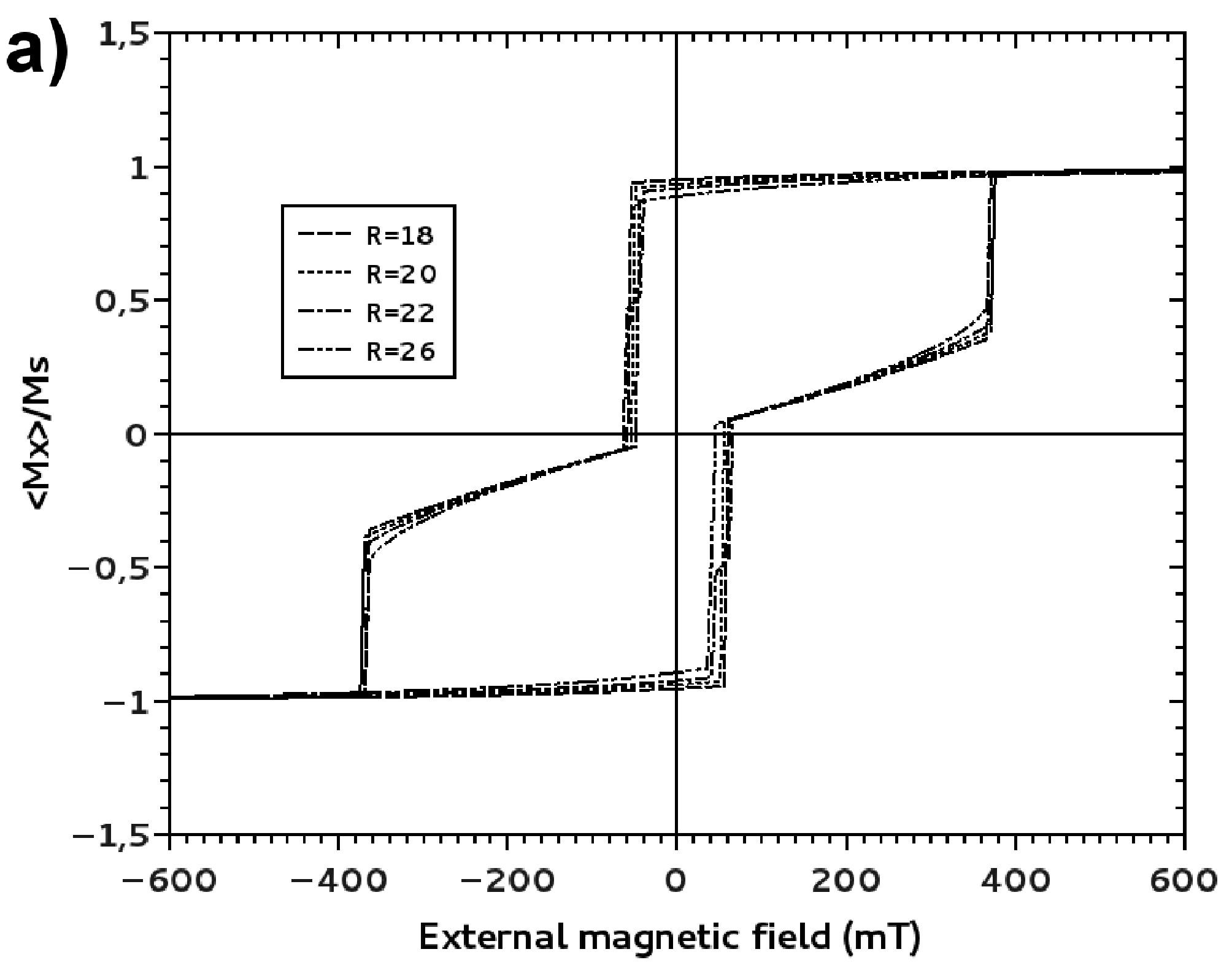}\includegraphics[scale=0.3]{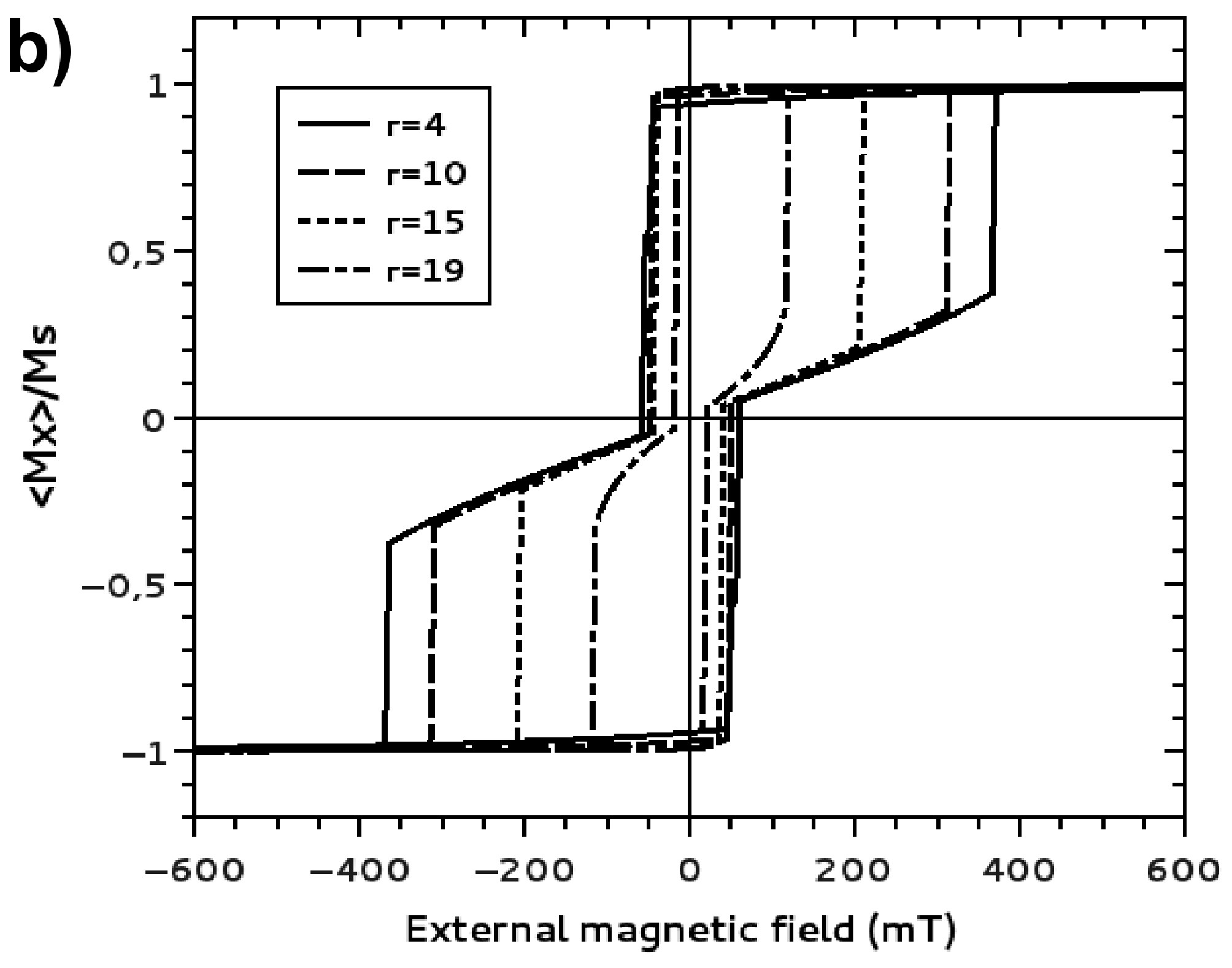}\caption{a) Hysteresis loops for magnetic nanotori with $r=4$ nm and large $R$ under an in-plane magnetic field. b) Hysteresis loops for $R =20$ nm and different values of $r$}\label{Hysteresis-20nm}
\end{figure}

\begin{figure}
\includegraphics[scale=1]{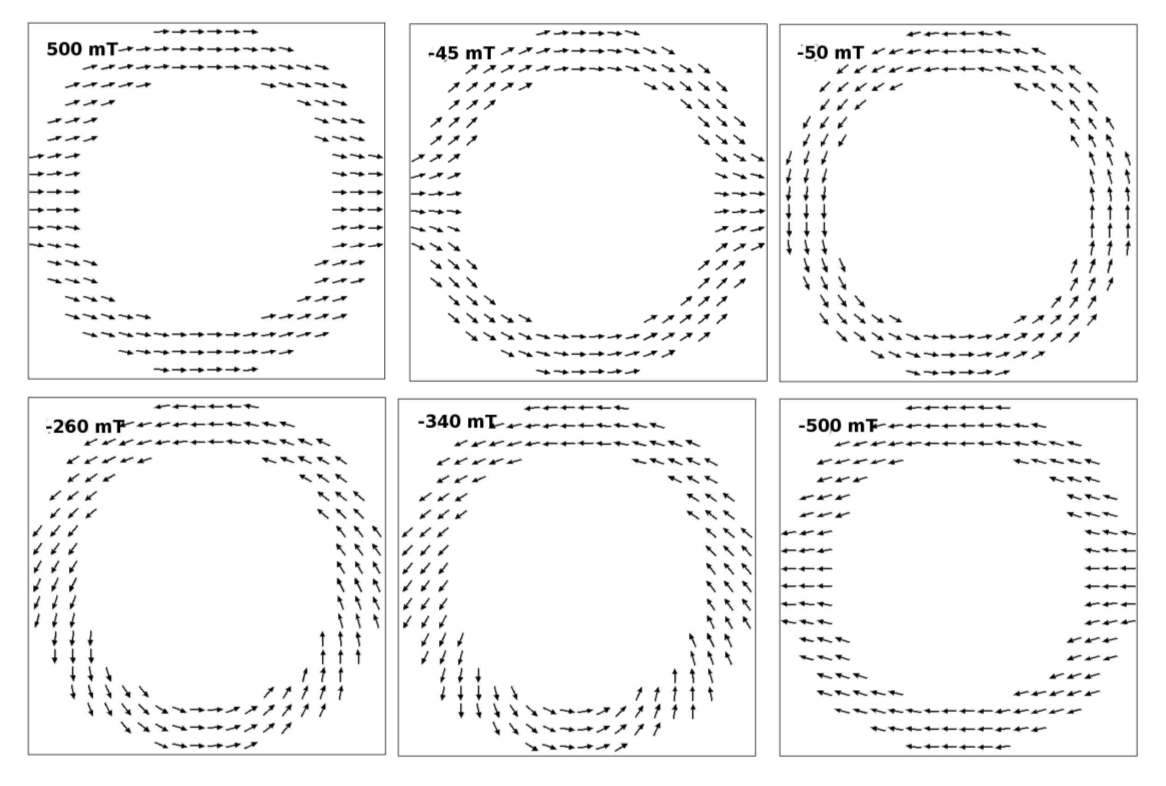}
\caption{Snapshots of the magnetization reversal process of a torus with $R=26$ nm and $r=4$ nm at different filed values}\label{Fig-x-thin}
\end{figure}

To understand this behavior we obtain snapshots of the magnetization for different values of the magnetic field for a nanotorus with $R = 10$ nm and $r = 4$ nm.  Our results are presented in Fig. 10. The full data of the reversal process is part of a video included as Supplementary material \cite{Supplemental-material}.  From the analysis of the video and Fig. 10 it can be observed that the square loops are result of a coherent and fast reversal of the magnetization. Since magnetostatic and exchange energies do not change during the rotation, a fast reversion occurs in order to decrease the larger Zeeman energy, resulting in a hysteresis curve with a square shape. The lower coercivities associated to fix $r$ and larger values of $R$ -and consequently larger magnetic volumes- are result of the need of a faster rotation of the magnetic moments to decrease the Zeeman energy. The whole process illustrated in Fig. 10 occurs at $H_{\text{x}}$ = -120 mT.

\begin{figure}
\includegraphics[scale=0.3]{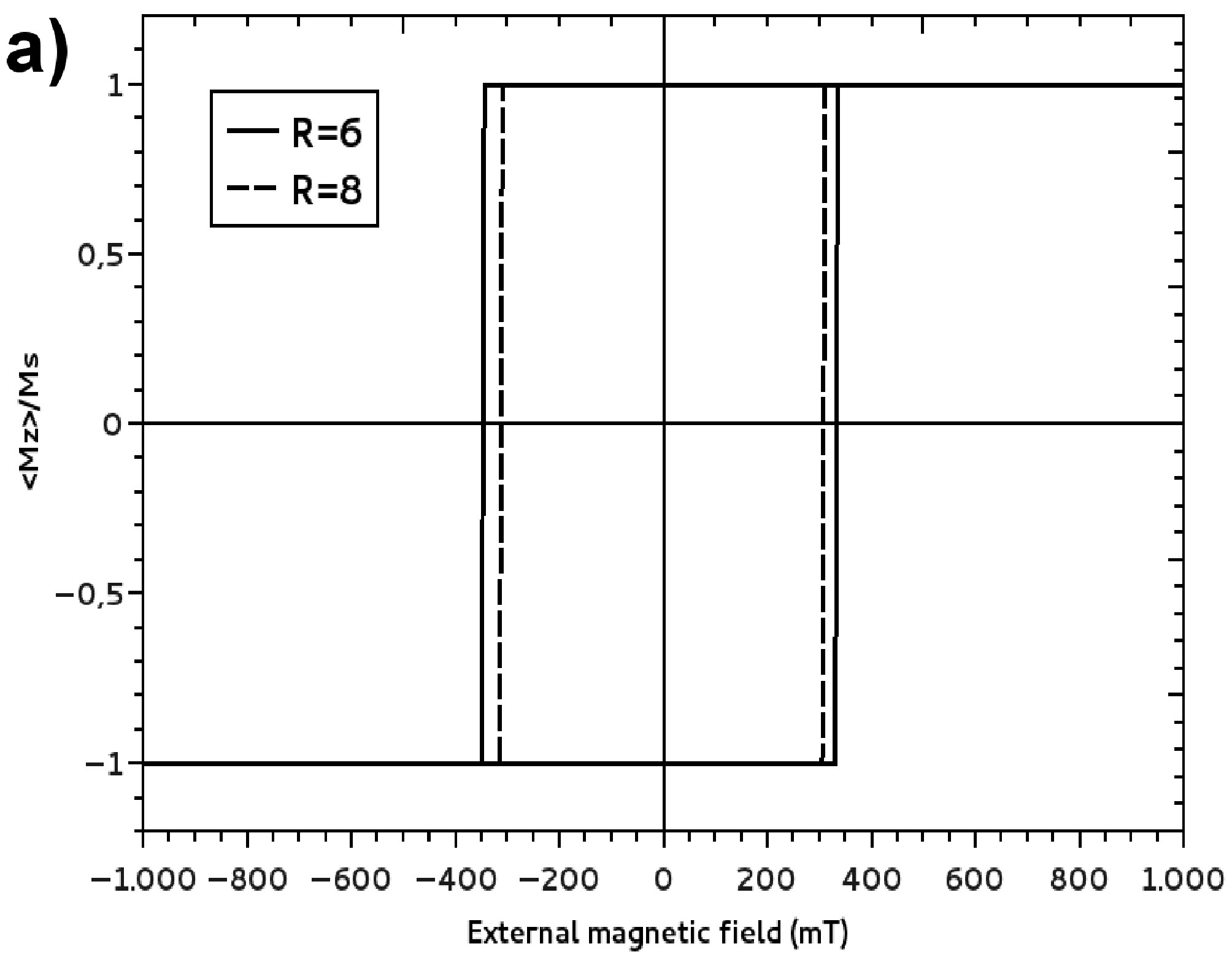}\includegraphics[scale=0.3]{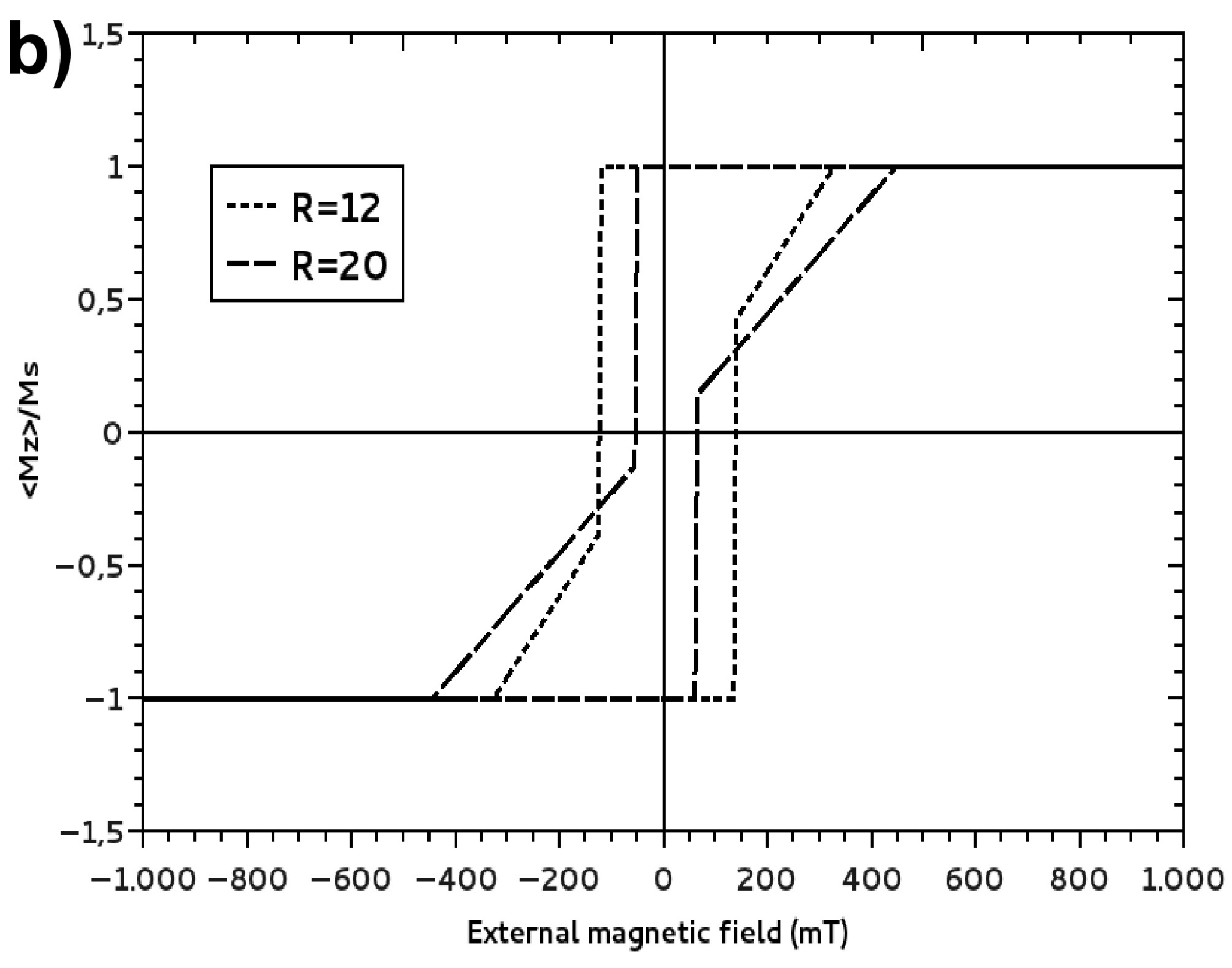}\caption{Hysteresis curves for toroidal nanomagnets with $r = 4$ nm and different values of $R$ for a magnetic field pointing along the $z$-axis.}\label{Hysteresis-z}
\end{figure}

\begin{figure}
\includegraphics[scale=0.68]{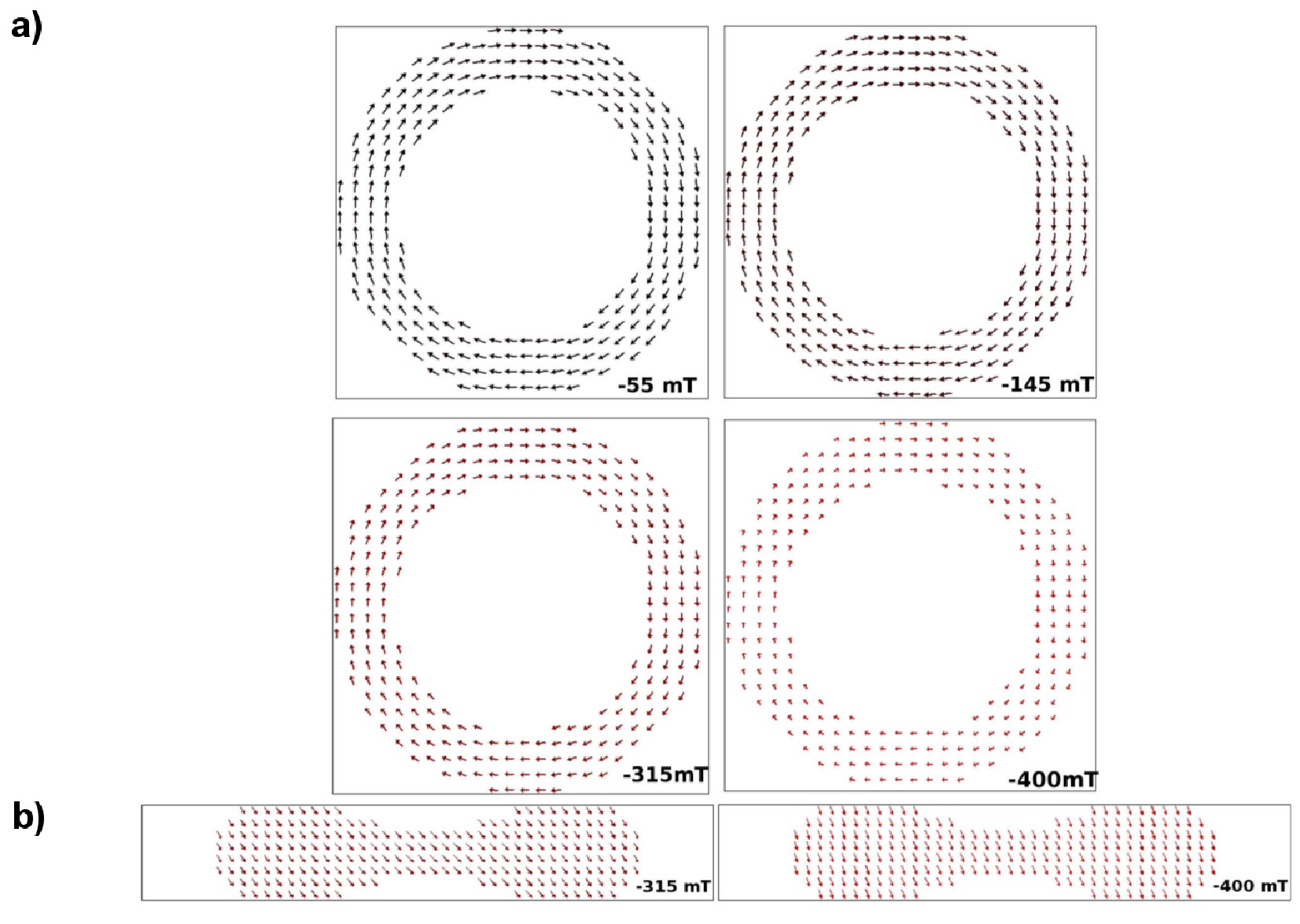}\caption{ Snapshots of the  $\mathcal{V}$ reversal mode under an out-of-plane magnetic field. It can be notice that a) a in-plane vortex appears when the magnetic field is $H_z = -55$ mT, while b) a coherent rotation is evidenced for the $z$ component of the magnetization.}\label{Fig2-z}
\end{figure}

We also consider large values of $R$, that is, $R \geq 18$ nm. Fig. 11a depict the hysteresis loops for nanotori with $r = 4$ nm and different $R$ while Fig. 11b presents the hysteresis loops for $R = 20$ nm, and $r$ varying from 4 to 19 nm. All these loops exhibit a clear neck with a coercivity and remanence  slightly depending on $r$. Snapshots of the reversal process for a torous of $R = 26$ nm and $r = 4$ nm are depicted in Fig. 12, while a video of the full process is included in the Supplementary material \cite{Supplemental-material}.  From the snapshots we observe that the neck is the result of the nucleation and propagation of a vortex. The reversion starts with the nucleation of an onion state that generates a vortex that propagates in a direction perpendicular to the field. As the field increases along the  $-x$-direction the vortex propagates  increasing the magnetization component along the  $-x$-direction in such a way that it seams to be deformed.  In this case there is a strong dependence of the annihilation field on the $r$ value. The shape of the hysteresis loop is due to the large vortex stability region in the phase diagram for $R \geq 18$ nm.

Finally we have performed micromagnetic simulations  for toroidal nanomagnets under an out-of-plane  magnetic field. Our results are illustrated in  Fig. \ref{Hysteresis-z} evidencing the same behavior characteristic of a coherent and a vortex reversal modes. That means, squared and with neck cycles.  However, in this case, the $\mathcal{V}$ reversal mode has some differences as compared to the in-plane field case. Such changes are understood  using the magnetization snapshots shown in Fig. \ref{Fig2-z}  for a nanotorus with $R=20$ nm and $r=4$ nm. It can be observed that the vortex nucleates at $H_z \sim-55 mT$. In this case the in-plane component of the magnetization forms a vortex, while the  out-of-plane component of the magnetization exhibits a coherent rotation. That is to say, during the reversal process, the magnetization configuration is well represented by $\mathbf{m}=m_\phi\hat{\phi}-\sqrt{1-m_\phi^2}\hat{z}$ where $m_{\phi}$ is the in plane component of the magnetization and decreases when the magnetic field increases in the $-z$-direction. At the annihilation field the vortex is smoothly deformed until all magnetic moments point along the $z$-direction. This smooth transition of the vortex to a SD$_\text{z}$ state occurs because a magnetic field pointing along the $z$-axis direction does not break the azimuthal symmetry of the vortex.

\section{Conclusions}
By performing micromagnetic simulations, we have obtained a phase diagram of a ferromagnetic nanotori. The magnetic energy of the final state associated to each one of the three possible configurations has been calculated, resulting in that a single domain state with the magnetization pointing along the out-of-plane direction is not possible for any geometrical parameters.  In addition, we have shown that as compared to  cylindrical rings, magnetic nanotori can support a vortex for smaller volumes. Therefore, if a vortex is proposed to be used as a bit of information, toroidal magnetic nanoparticles could generate more dense data storage devices. Finally, we have obtained hysteresis loops for toroidal nanomagnets for in-plane and out-of-plane external magnetic fields. In both cases, depending on the aspect ratio of the nanoparticle, $\mathcal{CR}$ or $\mathcal{V}$ reversal modes are observed.

In Chile we acknowledge financial support from FONDECYT 1120356 and 1150072, and Financiamiento Basal FB 0807 para Centros Cient\'ificos y Tecnol\'ogicos de Excelencia. We also acknowledge computational facilities from the Nanotechnology Laboratory at the University Diego Portales. In Brazil we acknowledge the financial support of CNPq.


\begin{thebibliography}{99}
\bibitem{Cowburn-JPD-33-2000}
R.P. Cowburn, J. Phys. \textbf{D 33}, R1 (2000).

\bibitem{Fabric-1} 
R. Streubel, V.P. Kravchuk, D.D. Sheka, D. Makarov, F. Kronast, O.G. Schmidt, and Y. Gaididei, Appl. Phys Lett. \textbf{101}, 132419 (2012).

\bibitem{Fabric-2} 
R. Streubel, D.J. Thurmer, D. Makarov, F. Kronast, T. Kosub, V.P. Kravchuk, D.D. Sheka, Y. Gaididei, R. Sch\"afer, and O.G. Schmidt, Nano Lett. \textbf{12}, 3961 (2012).

\bibitem{Fabric-3} 
I. Minguez-Bacho, S. Rodriguez-L\'opez, M. V\'azquez, M. Hern\'andez-V\'elez, and K. Nielsch, Nanotechnology \textbf{25}, 145301 (2014).

\bibitem{Fabric-4}
A.P. Safronov, I.V. Beketov, S.V. Komogortsev, G.V. Kurlyandskaya, A.I. Medvedev, D.V. Leiman, A. Larrañaga, and S.M. Bhagat, AIP Advances \textbf{3}, 052135 (2013).

\bibitem{Fabric-5}
R. Streubel, F. Kronast, P. Fischer, D. Parkinson, O.G. Schimidt, and D. Makarov, Nat. Commun. \textbf{6}, 7612 (2015).

\bibitem{Possibilities}
J. Dai, J.-Q. Wang, C. Sangregorio, J. Fang, E. Carpenter, and J. Tang, J.
Appl. Phys. \textbf{87}, 7397 (2000); S. H. Sun, C. B. Murray, D. Weller, L. Folks, and A. Moser, Science \textbf{287}, 1989 (2000); R. Hertel, Nat. Nanotechnol. \textbf{8}, 318 (2013).

\bibitem{Cancer-therapy}
D.-H. Kim, E. Rozhkova, I. Ulasov, S. Bader, T. Rajh, M. Lesniak, and V.
Novosad, Nature Mater. \textbf{9}, 165 (2009).

\bibitem{Goll-PRB-2004}
D. Goll, A.E. Berkowitz, and H.N. Bertram, Phys. Rev. \textbf{B 70}, 184432 (2004); B. Barpanda, T. Kasama, R.E. Dunin-Borkowski, M.R. Sheinfein, and A.S. Arrot, J. Appl. Phys. \textbf{99}, 08G103 (2006).

\bibitem{Russier-JAP-2009}
V. Russier, J. Appl. Phys. \textbf{105}, 073915 (2009)

\bibitem{Kravchuk-PRB-2012}
V.P. Kravchuk, D.D. Sheka, R. Streubel, D. Makarov, O.G. Schmidt, and Y. Gaididei, Phys. Rev. \textbf{B 85}, 144433 (2012).

\bibitem{Landeros-JAP}
P. Landeros, and \'A.S. N\'u\~nez, J. Appl. Phys. \textbf{108}, 033917 (2010); A. Gonz\'alez, P. Landeros, and \'A.S. N\'u\~nez, J. Magn. Mag. Mat. \textbf{322}, 530 (2010).

\bibitem{Landeros-APL-2007}
P. Landeros, S. Allende, J. Escrig, E. Salcedo, D. Altbir, and E.E. Vogel, Appl. Phys. Lett. \textbf{90}, 102501 (2007); R. Wieser, E.Y. Vedmedenko, P. Weinberger, and R. Wiesendanger, Phys. Rev. \textbf{B 82}, 144430 (2010).

\bibitem{Cone-paper}
W.A. Freitas, W.A. Moura-Melo, A.R. Pereira, Phys. Lett. \textbf{A 336}, 412 (2005).

\bibitem{Gaididei-PRL-2014}
Y. Gaididei, V.P. Kravchuk, and D.D. Sheka, Phys. Rev. Lett. \textbf{112}, 257203 (2014).

\bibitem{Sheka-JPA-2014}
D.D. Sheka, V.P. Kravchuk, and Y. Gaididei, J. Phys. \textbf{A 48}, 125202 (2015).

\bibitem{Kravchuk-Mobius}
O.V. Pylypovskyi, V.P. Kravchuk, D.D. Sheka, D. Makarov, O.G. Schmidt, and Y. Gaididei, Phys. Rev. Lett. \textbf{114}, 197204 (2015).

\bibitem{Kravchuk-Helical}
D.D. Sheka, V.P. Kravchuk, K.V. Yershov, and Yuri Gaididei, Phys. Rev. \textbf{B 92}, 054417 (2015).

\bibitem{Kravchuk-curved-nanowire}
K.V. Yershov, V.P. Kravchuk, D.D. Sheka, and Y. Gaididei, Phys. Rev. \textbf{B 92},
104412 (2015).

\bibitem{Vagson-JAP-2015}
V.L. Carvalho-Santos, R.G. Elias, J.M. Fonseca, and D. Altbir, J. Appl. Phys. \textbf{117}, 17E518 (2015).

\bibitem{Priscila-PLA-2015}
P.S.C. Vilas-Boas, R.G. Elias, D. Altbir, J.M. Fonseca, and V.L. Carvalho-Santos, Phys. Lett. \textbf{A 379}, 47 (2015).

\bibitem{Vagson-JMMM-2015}
V.L. Carvalho-Santos, R.G. Elias, D. Altbir, J.M. Fonseca, J. Magn. Mag Mat. \textbf{391}, 179 (2015).

\bibitem{Klaui-works}
M. Kl\"aui, C.A.F. Vaz, L. Lopez-Diaz, and J.A.C. Bland, J. Phys.: Condens. Matter \textbf{15}, R985 (2003).

\bibitem{Yoo-APL-2003}
Y.G. Yoo, M. Kl\"aui, C.A.F. Vaz, L.J. Heyderman, and J.A.C. Bland, Appl. Phys. Lett. \textbf{82}, 2470 (2003).

\bibitem{Heyderman-JAP-2003}
L.J. Heyderman, C. David, M. Kl\"aui, C.A.F. Vaz, and J.A.C. Bland, J. Appl. Phys. \textbf{93}, 10011 (2003).

\bibitem{Klaui-JAP-2004}
M. Kl\"aui, C.A.F. Vaz, J.A.C. Bland, L.J. Heyderman, C. David, E.H.P. Sinnecker, and A.P. Guimar\~aes, J. Appl. Phys. \textbf{95}, 6639 (2004).

\bibitem{Klaui-APL-2004}
M. Kl\"aui, C.A.F. Vaz, J.A.C. Bland, E.H.P. Sinnecker, and A.P. Guimar\~aes, W. Wernsdorfer, G Faini, E. Cambrill, L.J. Heyderman, and C. David, Appl. Phys. Lett. \textbf{84}, 951 (2004).

\bibitem{Bellegia-JMMM-2006}
M. Beleggia, J. W. Lau, M. A. Schofield, Y. Zhu, S. Tandon, and M. De
Graef, J. Magn. Magn. Mater. \textbf{301}, 131 (2006).

\bibitem{Ross-JAP-2006}
C.A. Ross, F.J. Casta\~no, D. Morecroft, W. Jung, H.I. Smith, T.A. Moore, T.J. Hayward, J.A.C. Bland, T.J. Bromwich, and A.K. Petford-Long, J. Appl. Phys. \textbf{99}, 08S501 (2006).

\bibitem{Yang-APL-2007}
T. Yang, M. Hara, A. Hirohata, T. Kimura, and Y. Otani, Appl. Phys. Lett. \textbf{90}, 022504 (2007).

\bibitem{Zhu-PRL-2006}
F. Q. Zhu, G. W. Chern, O. Tchernyshyov, X. C. Zhu, J. G. Zhu, and C. L.
Chien, Phys. Rev. Lett. \textbf{96}, 027205 (2006).

\bibitem{Vaz-JPC-2007}
C. A. F. Vaz, T. J. Hayward, J. Llandro, F. Schackert, D. Morecroft, J. A. C. Bland, M. Kl\"aui, M. Laufenberg, D. Backes, U. R\"udiger, F. J. Casta\~no, C. A. Ross, L. J. Heyderman, F. Nolting, A. Locatelli, G. Faini, S. Cherifi, and W. Wensdorfer, J. Phys.: Condens. Matter \textbf{19}, 255207 (2007).

\bibitem{Kravchuck-nanoring}
V. P. Kravchuk, D. D. Sheka, and Y. B. Gaididei, J. Magn. Magn. Mater.
\textbf{310}, 116 (2007).

\bibitem{Castillo-works}
S. Castillo-Sep\'ulveda, N. M. Vargas, R.A. Escobar, S.E. Baltazar, S. Allende, and D. Altbir, J. Appl. Phys. \textbf{115}, 223903 (2014).

\bibitem{Carbon-nanotori}
E. Yazgan, E. Ta\c sci, O. B. Malcio\u glu, and \c S. Erko\c c, J. Mol. Struct. \textbf{681}, 231 (2004); E. Ta\c sci, E. Yazgan, O. B. Malcio\u glu, and \c S. Erko\c c, Fullerenes, Nanotubes, Carbon Nanostruct. \textbf{13}, 147 (2005).

\bibitem{polymer-tori}
J. Lee,	K. Baek, M. Kim,	G. Yun,	Y.H. Ko, N.-S. Lee,	I. Hwang, J. Kim, R. Natarajan, C.G. Park, W. Sung, and K. Kim, Nature Chemistry \textbf{6}, 97 (2014).

\bibitem{Vagson-JAP-2010}
V.L. Carvalho-Santos, W.A. Moura-Melo, and A.R. Pereira, J. Appl. Phys. \textbf{108}, 094310 (2010).

\bibitem{Bellegia-Torus}
M. Bellegia, M. de Graef, and Y.T. Millev, Proc. R. Soc. \textbf{A 465} 3581 (2009).

\bibitem{Dandoloff-torus}
A. Saxena, R. Dandoloff, and T. Lookman, Physica \textbf{A 261}, 13 (1998).

\bibitem{Vagson-PRB-2008}
V.L. Carvalho-Santos, A.R. Moura, W.A. Moura-Melo, A.R. Pereira, Phys. Rev. \textbf{B 77}, 134450 (2008).

\bibitem{oommf-code}
M.J. Donahue, and D.G. Porter, National Institute of Standards and Technology Interagency Reports NISTIR \textbf{6376} (1999).


\bibitem{Landau-paper}
L. Landau, and E. Lifshitz, Ukr. J. Phys. \textbf{53}, 14 (2008). Reprinted from Phys. Zeitsch. der Sow. \textbf{8}, 153 (1935).


\bibitem{Gilbert-paper}
T.L. Gilbert, IEEE Trans. on Magnetics \textbf{40}, 3443 (2004).

\bibitem{Landeros-JAP-2006}
P. Landeros, J. Escrig, D. Altbir, M. Bahiana, and J. d’Albuquerque e
Castro, J. Appl. Phys. \textbf{100}, 044311 (2006).

\bibitem{Supplemental-material}
Supplemental Materials, link to be given 

\end{thebibliography}
\end{document}